\documentclass[floatfix,prd,twocolumn,letterpaper,lengthcheck,superscriptaddress,showpacs,amssymb,amsmath,amsfonts,aps,altaffilletter,nofootinbib,nopreprintnumbers,showpacs,longbibliography]{revtex4-1}

\usepackage{color}
\usepackage{hyperref}
\usepackage{amsmath}
\usepackage{multirow}
\usepackage{graphicx}
\usepackage{caption}
\usepackage{subcaption}
\usepackage[colorinlistoftodos]{todonotes}
\usepackage{float}

\begin{document}

\newcommand{\rhat}{\hat{r}}
\newcommand{\iotahat}{\hat{\iota}}
\newcommand{\phihat}{\hat{\phi}}
\newcommand{\h}{\mathfrak{h}}
\newcommand{\vek}[1]{\boldsymbol{#1}}
\newcommand{\IUCAA}{\affiliation{Inter-University Centre for Astronomy and Astrophysics, Post Bag 4, Ganeshkhind, Pune 411 007, India}}
\newcommand{\WSU}{\affiliation{Department of Physics \& Astronomy, Washington State University, 1245 Webster, Pullman, WA 99164-2814, U.S.A}}
\newcommand{\AEI}{\affiliation{Max Planck Institute for Gravitational Physics (Albert Einstein Institute), Hannover, Germany}} 
    
\author{Vaishak Prasad}
\affiliation{Inter-University Centre for Astronomy and Astrophysics, Post Bag 4, Ganeshkhind, Pune 411 007, India}

\author{Anshu Gupta}
\affiliation{Inter-University Centre for Astronomy and Astrophysics, Post Bag 4, Ganeshkhind, Pune 411 007, India}

\author{Sukanta Bose}
\affiliation{Inter-University Centre for Astronomy and Astrophysics, Post Bag 4, Ganeshkhind, Pune 411 007, India}
\affiliation{Department of Physics and Astronomy, Washington State University, 1245 Webster, Pullman, WA 99164-2814, U.S.A}

\author{Badri Krishnan}
\affiliation{Max-Planck-Institut f\"ur Gravitationsphysik (Albert Einstein Institute), Callinstr. 38, 30167 Hannover, Germany}
\affiliation{Leibniz  Universit\"at  Hannover,  Welfengarten  1-A,  D-30167  Hannover,  Germany}
\affiliation{Institute for Mathematics, Astrophysics and Particle Physics, Radboud University, Heyendaalseweg 135, 6525 AJ Nijmegen, The Netherlands}

\title[]{Tidal deformation of dynamical horizons in binary black hole mergers}

\date{\today}

\begin{abstract}

An important physical phenomenon that manifests itself during the
inspiral of two orbiting compact objects is the tidal deformation of each under the
gravitational influence of its companion.  In the case of binary neutron
star mergers, this tidal deformation and the associated Love numbers
have been used to probe properties of dense matter and the nuclear
equation of state.  Non-spinning black holes on the other hand have a vanishing
(field) tidal Love number in General Relativity.  
This pertains to the deformation of the asymptotic gravitational
field.  In certain cases, especially in the late stages of the
inspiral phase when the black holes get close to each other, the
\emph{source} multipole moments might be more relevant in probing
their properties and the No-Hair
conjecture; contrastingly, these Love
numbers do \emph{not} vanish. In this paper, we track the source
multipole moments in simulations of several binary black hole mergers and
calculate these Love numbers.  We present evidence that, at least for
modest mass ratios, the behavior of the source multipole moments is
universal.  [{\color{blue} This manuscript has been assigned the LIGO Preprint number LIGO-P2100109.}]

\end{abstract}

\maketitle

\section{Introduction}
\label{sec:intro}

In this work, we study the relation between the geometry of the dynamical horizons and their tidal environment in a binary black hole scenario. We address the problem of tidal deformability of black holes in the strong field regime.

The general treatment of the tidal deformation of compact objects has been perturbative. In such a treatment, an external tidal field is taken to induce a deformation of the compact object, and consequently the potential or the asymptotic configuration of the gravitational field sourced by the compact object also changes \cite{binningtonpoisson2009}. In the non-Relativistic theory, the deformation of the compact object can be quantified using its mass multipole moments, which are expressed as integrals over the source distribution. The external tidal field deforms the compact object (the source) and changes its distribution, i.e., the multipole moments.  At the linear order in the external tidal field, every $l^{\rm th}$ multipolar order of the external tidal field induces an $l^{\rm th}$ multipolar moment in the configuration of the compact object, and the constant of proportionality relating these is called a Love number. These Love numbers encode information about the constituent materials of the source. 

In General Relativity, the situation is more complex. The deformation of the compact object is now conveniently quantified using two sets of multipole moments: mass and current type multipole moments, which quantify the configuration of the asymptotic gravitational field produced by the source. Although Love numbers can again be described as linearly relating the change in the field multipole moments and the external tidal field, the relation is not restricted to a simple proportionality. When the source object is spinning, the relation is tensorial and, hence, can involve the mixing of tidal fields at different multipolar orders. Thus, when the compact object is spinning, the $l^{\rm th}$ order multipole moment induced in the asymptotic field produced by the compact object can have contributions from any $l^{'{\rm th}}$ order component of the external tidal field with $l\neq l'$.

A sizeable literature exists on applications of tidal deformation and Love numbers~\cite{1955MNRAS.115..101B,Flanagan:2007ix,Damour:2009vw}, especially, in how they encode information about neutron star equation of state~\cite{Lattimer:2015nhk} in gravitational-wave signals~\cite{Abbott:2018exr,De:2018uhw,Capano:2019eae}.
On the other hand, tidal deformations of black holes and the associated Love numbers are discussed in, e.g., Refs.~\cite{Landry2014,Damour:2009va,binningtonpoisson2009} for the non-spinning case and Refs.~\cite{Pani:2015hfa,LeTiec:2020bos,LeTiec:2020spy} for the spinning case.

Since a black hole is a solution to vacuum field equations, the usual definition of the multipolar configuration of the system involving an integral of 
its mass density is irrelevant here. Furthermore, during a dynamical scenario such as a binary black hole merger, which is of great relevance to the recent gravitational-wave detections, it is difficult to define and study the change in the asymptotic gravitational field of each black hole as it inspirals toward the other and merge. Also, the situation is highly dynamical
as the two objects orbit under the mutual influence of their strong field gravity. 
Thus, the field Love numbers will pose a limitation to understanding this problem in the strong field regime.

Typically, past treatments of tidal deformability either start with the assumption that there exists a source distribution that is deformed due to an external tidal field \cite{DamourNagar2009} or involve the application of black hole perturbation theory~\cite{binningtonpoisson2009, Damour:2009va, Landry2014, Pani:2015hfa, LeTiec:2020bos, LeTiec:2020spy}. In the former treatment, a limit of the compactness is then taken to deduce the corresponding Love numbers for black holes. The problem with the ``limiting compactness" approach is that black holes do not have a source distribution in the same sense as that of, e.g., a neutron star. Furthermore, in a true sense, the horizons of black holes evolve, cease to be null hypersurfaces and are no longer isolated in a dynamical tidal environment.

In the perturbative approach, the gravitational field far away from the blackhole is studied. Often, simplifying assumptions, such as staticity / stationarity, slow variation in time, etc., are imposed. The perturbative approaches cannot be used to understand (a) the deformation of black holes when they are close to each other, and (b) the fully dynamical aspects of tides on black holes. Furthermore, in a numerical simulation, one does not have access to the asymptotic regions of the individual black holes to study the problem dynamically using the gravitational field far away from the system. Due to these reasons, it is desirable to employ an alternative treatment with a differently defined set of quantities that characterize the tidal deformability of black holes in the strong field regime.

Hence, more appropriately, one should address how the external tidal field induces a change in the geometry of the horizon. One has to study the tidal deformability of the quasi-local \emph{dynamical horizons} rather than the teleological event horizons, which only coincide with the former in the special case of stationarity. The information on the geometry of the dynamical horizon is encoded in its source multipole moments and can be used to study the problem of tidal deformability. One may expect a similar relation between the source multipoles and the $l^{\rm th}$ order multipolar external tidal field. In this sense, the dynamical horizon formalism is an ideally suited framework to treat the problem of the tidal deformation of black holes.

In this formalism, for black holes in a vacuum, a set of numbers -- called the source multipole moments -- fully characterize the horizon geometry.
In a tidal environment, these numbers can change, e.g., owing to an alteration in the horizon geometry in response to the external tidal field. Using this approach, one can directly study the deformation of the horizon geometry due to an external tidal field.
In this work, we address how the black holes are tidally deformed in a binary black hole merger scenario. We describe and use a convenient definition of tidal deformability suited to a binary black hole system and compute them numerically using full numerical relativity simulation data of binary black hole mergers of non-spinning black holes.
We use the framework of quasi-local horizons for our analysis; see e.g. \cite{Ashtekar:2004cn,Booth:2005qc} for reviews. We use the source multipole moments of the involved dynamical horizons to address the problem of their tidal deformability.  In particular, we use the mass multipole moments to characterize the deformation in the horizon geometry of non-spinning black holes and define a set of dimensionless tidal coefficients. We then calculate the leading order tidal coefficients that characterize the deformation of a black hole's multipole moments in the tidal environment of its companion in a binary black hole merger scenario. 

The plan for the rest of the paper is as follows. Sec.~\ref{sec:prelim} briefly reviews basic concepts and equations for dynamical horizons and describes the definition of the source multipole moments of horizons. The details of Numerical simulations are given in Sec.~\ref{sec:nr}. In Sec. ~\ref{sec:com_tidal_coeffs} we compute the tidal coefficients and describe the fitting procedure. The results are summarised in Sec.~\ref{sec:results} followed by Sec. ~\ref{sec:conclusions} with discussion and conclusions. Appendix~\ref{sec:appendix} contains comparison of distance measures, among the centroid distance and various orders of post-Newtonian approximations.

All the equations and quantities are expressed in geometric units, where $G = c = 1$. The masses of the horizons of the primary and secondary black holes in the initial data are $M_1$ and $M_2$, and their mass ratio is defined to be $q \equiv M_2/ M_1$, which is always $\leq 1$. Additionally, the total mass of the black hole horizons in the initial data, denoted by $M = M_1 + M_2$ is set to one. In a few places where tracking the dependence on $M$ is important, we show it explicitly. 

\section{Preliminaries}
\label{sec:prelim} 

Our calculation of the tidal deformations of black holes is based on the formalism of quasi-local horizons~\cite{Hayward:2000ca,Booth:2005qc,Ashtekar:2004cn,Gourgoulhon:2005ng,Visser:2009xp,Jaramillo:2011zw,Faraoni:2015pmn}. 
A detailed description of this formalism is beyond the scope of this article and we shall restrict ourselves to a brief overview of the most relevant concept, namely, that of black hole source multipole moments. 

The starting point is the notion of a marginally trapped surface, first introduced by Penrose in the context of the black hole singularity theorems \cite{Penrose:1964wq}. More specifically we need here the notion of marginally outer trapped surfaces (MOTS), which are closed space-like 2-dimensional surfaces of spherical topology, such that their outgoing null-normals $\ell^a$ have vanishing expansion $\Theta_{(\ell)}$. Thus, if $\mathcal{S}$ is a MOTS, $\Tilde{q}_{ab}$ the Riemannian metric on $\mathcal{S}$, and $\ell^a$ an outward-pointing null-normal to $\mathcal{S}$, then a MOTS has 
\begin{equation}
  \Theta_{(\ell)} := \Tilde{q}^{ab}\nabla_a\ell_b = 0\,.
\end{equation}
Under time evolution, a MOTS $\mathcal{S}$ traces out a 3-dimensional world tube sometimes referred to as a dynamical horizon \cite{Pook-Kolb:2020zhm,Pook-Kolb:2020jlr}, or a marginally trapped tube \cite{Andersson:2008up,Andersson:2007fh}. We shall not delve here into properties of this time evolution. We shall instead just consider various properties of $\mathcal{S}$ as functions of time. It turns out that the time evolution of MOTS is generally found to be smooth, which means that we end up with smooth functions of time. When two black holes collide, the process by how two distinct dynamical horizons merge to yield a single final dynamical horizon turns out to have interesting topological and dynamical properties \cite{PhysRevD.100.084044,PhysRevLett.123.171102,Pook-Kolb:2021gsh,Booth:2021sow,Booth:2021sow}. Again, this is beyond the scope of this paper. Here, we shall only consider the two dynamical horizons corresponding to the two individual black holes before they merge. 

For each of the two dynamical horizons, we shall calculate the source multipole moments. These were first introduced in \cite{Ashtekar:2004gp} for isolated horizons, and extended and used in \cite{Schnetter:2006yt} for dynamical horizons. These multipole moments have found applications, for example, in predictions of the anti-kick in binary black hole mergers \cite{Rezzolla:2010df} and for studying tidal deformations of black holes \cite{Cabero:2014nza,Gurlebeck:2015xpa}.

For defining these multipole moments, let $\mathcal{S}$ be a MOTS on a Cauchy surface $\Sigma$. Let $K_{ab}$ be the extrinsic curvature of $\Sigma$ embedded in spacetime, $r^a$ the unit space-like normal to $\mathcal{S}$ and tangent to $\Sigma$. Let $\mathcal{S}$ have the 2-Ricci scalar $\Tilde{\mathcal{R}}$. Also, let its areal radius be denoted by $R_{\mathcal{S}}$, its mass by $M_\mathcal{S}$ and its angular momentum by $J_S$. Furthermore, let $\mathcal{S}$ be axisymmetric and let $\varphi^a$ be the axial symmetry vector field on $\mathcal{S}$.  The symmetry vector $\varphi^a$ can be used to construct a preferred coordinate system $(\theta,\varphi)$ on $\mathcal{S}$ analogous to the usual spherical coordinates on a sphere; let $\zeta = \cos\theta$. We can then use spherical harmonics in this preferred coordinate system to construct multipole moments.  As expected we have two sets of moments $M_n$, $J_n$ such that $M_0$ is the mass $M_{\mathcal{S}}$ and $J_1$ is the angular momentum $J_{\mathcal{S}}$.  The expressions for the multipole moments  are the following
\begin{equation}
  \mathcal{M}_n = \frac{M_{\mathcal{S}}R_{\mathcal{S}}^n}{8\pi}\oint_S \Tilde{\mathcal{R}} P_n(\zeta) d^2S\,, \label{mass_mult}
\end{equation}
and 
\begin{equation}
  \mathcal{J}_n = \frac{R_{\mathcal{S}}^{n-1}}{8\pi} \oint_S P_n^\prime(\zeta) K_{ab}\varphi^a r^b d^2S \,, \label{spin_mult}
\end{equation}
where $P_n(\zeta)$ is the $n^{\mathrm{th}}$ Legendre polynomial and $P_n^\prime(\zeta)$ its derivative.

In a binary black hole merger scenario, the individual horizons of the black holes are dynamical and many of the above assumptions do not hold. For example, the Weyl scalar $\Psi_2$ of the two black holes are not time-independent, and neither are their areas, curvature, and other geometric quantities on $\mathcal{S}$. However, following Ref.~\cite{Schnetter:2006yt}, we shall continue to interpret the surface density and current in the same way so that the multipole moments share the same definitions as above.

The dynamical horizons are geometric objects that exist in spacetime independent of the spacetime foliation used to locate them. In this sense, the dynamical horizons are gauge invariant. On the other hand, the computation of the multipole moments in a numerical relativity simulation requires choosing a spacetime foliation. A different choice of spacetime slicing will give different values of the multipole moments $(\mathcal{M}_n, \mathcal{J}_n)$. For every choice of slicing the physical laws, like the flux/balance laws, will hold for  the corresponding dynamical horizon. The foliations and gauges we choose (1+log slicing, $\Gamma$-driver shift \cite{alcubierre2003h}) are such that the horizon geometries are close to Kerr when the black holes are far apart (in the start of the simulation)  and they settle down close to Kerr at late times after the merger. In all numerical  simulations done so far using these gauge choices, we find the same reasonable behaviour as expected \cite{alex2011}, just as in the case of waveform extraction. Although bad gauge choices are expected to affect the multipole moment values, we do not expect them to change significantly when the gauge choice is reasonable. We expect that our results will be the same qualitatively for other reasonable choices of gauge. A quantitative study of the dependence of these results on the choice of gauge is beyond the scope of this work.

In this work, we attempt to understand the tidal deformation of the horizons using the dynamical horizon formalism. The horizon geometry is characterized by the mass and spin multipole moments of the dynamical horizon. The presence of the companion induces a change in the horizon geometry of the black holes. For instance, the 2-Ricci scalars ($\Tilde{\mathcal{R}}$) of the two-dimensional slices of the dynamical horizons change from their equilibrium configuration,  i.e., from their isolated horizon values. During the inspiral phase, the two black holes influence each other and mutually change their horizon geometries. As they inspiral toward each other in the tidal environment of their companion the source multipole moments (as defined in Eqs.~\eqref{mass_mult} and \eqref{spin_mult}), which encode information about the deformed geometries of the MOTS of the individual dynamical horizons, evolve in time.

In this work, as we deal with non-spinning black holes, we attempt to understand the tidal deformability of black holes using the mass multipole moments. Using the evolution of the mass multipole moment of the MOTS of the dynamical horizons of both the black holes, we define and compute dimensionless tidal coefficients (Love numbers)~\footnote{We use \emph{tidal coefficients} and \emph{Love numbers} interchangeably} that universally characterize the tidal deformability of black holes in General Relativity. Although our methods are generically applicable to all multipolar deformations, in this work we will study tidal deformation of black holes using $\mathcal{M}_2$ as defined in Eq.~\eqref{mass_mult}.

Turning now to our object of interest, namely a binary black hole system in the inspiral/pre-merger phase. Imagine starting with two Kerr black holes far apart, with the source mass and spin multipoles $(\mathcal{M}_n,\mathcal{J}_n)$ defined above exactly as for a Kerr black hole. As the black holes spiral in and approach each other, these multipole moments will vary under the influence of the gravitational field of the other. Let the variation in $\mathcal{M}_n$ and $\mathcal{J}_n$ be respectively $\delta\mathcal{M}_n$ and $\delta\mathcal{J}_n$ (these will be functions of time). In general, this deformation of the multipole moments will depend on the mass and spin of the companion, the separation $d$ between the black holes, and also possibly their relative velocity. However, for simplicity, we will not study the dependence on all these parameters in this work. The relative velocity would be important in describing the evolution of the multipole moments in the later stages of the inspiral and the merger, and is not included here. Furthermore, we do not study the effect due to spinning black holes.

Note that as the horizon masses of the two black holes evolve during the inspiral, they deviate from the physical masses $M_{1,2}$ in the initial data. However, the fractional deviation turns out to be sub-percent in the analysis domain (i.e., $\lessapprox 0.1$\%). Similarly, the least upper bound for the spins acquired by the black holes was found to be $\lessapprox 10^{-2}$.

We assume the masses and spins to remain constant throughout the evolution. Therefore, in what follows, we restrict ourselves to the simpler situation of non-spinning black holes, and also ignore their relative velocity.

We require that in the limit $d\rightarrow \infty$, $\delta\mathcal{M}_n$ and $\delta\mathcal{J}_n$ should vanish. From dimensional arguments, we are then naturally led to an expansion of the multipole moment of the black hole whose deformation is being studied. Denoting its mass by $M_{td}$ and that of its companion responsible for the tidal field by $M_{tf}$, the expansion can be written in the form:
\begin{align}
  \dfrac{\delta \mathcal{M}_n}{M_{td}^{n+1}} &= \sum_{i,j = 1}^{\infty} \alpha_{ij}^{(n)}\dfrac{M_{td}^i M_{tf}^j}{d^{i+j}}\,.
  \label{exp}
\end{align}
The deformation of the spin multipole moments $\mathcal{J}_n$ can also be expanded in a similar fashion. However, since we are dealing with non-spinning blackholes in this work, we will restrict ourselves to the behaviour of mass multipole moments.Such an expansion has been previously used to understand the tidal deformations of black holes in the Bowen-York initial data set \cite{Cabero:2014nza}. Along similar lines, the following observations can be made on the above expansion. The expansion must start at $i = j =1$ since when we take the limit of either of the masses tending to zero, the left-hand side must vanish. Also, there should be no positive exponents on the distance measure $d$; i.e., $\delta\mathcal{M}$ not contain any terms that are proportional to $d^{\nu > 0}$ since the perturbation should vanish in the limit $d\rightarrow \infty$. The coefficients $\alpha_{ij}^{(n)}$ are dimensionless coefficients independent of the parameters of the system as they have been scaled out by appropriate factors. These coefficients are therefore the same for all black holes and are thus universal. They should characterize the tidal deformability of the black holes. 

\section{Numerical simulations of binary black hole mergers}
\label{sec:nr}

We ran numerical simulations of orbital mergers for a set of non-spinning binary black holes with varying mass ratios. The systems are evolved numerically using puncture data \cite{PhysRevLett.78.3606} describing two black holes in quasi-circular orbits, with typical initial separations of $10$-$11 M$. These simulations cover the dynamical behavior starting at 5-6 orbits before the merger and going up to the merger phase. The gravitational waveform is extracted \cite{Baker:2002qf} at various distances between $100 M$ to $500 M$ from the merger location.

Since this work is focused on the study of individual horizon geometry of the black holes in the inspiral phase, we track the individual horizons up to merger and compute quasi-local quantities \cite{Dreyer:2002mx, Schnetter:2006yt} using the isolated and dynamical horizon formalism \cite{Ashtekar:2004cn}. 

Simulations are performed using the publicly available code Einstein Toolkit \cite{Loffler:2011ay, EinsteinToolkit:web}. The initial data is generated based on the puncture approach \cite{Ansorg:2004ds}, which has been evolved through BSSNOK formulation \cite{Alcubierre:2000xu, Alcubierre:2002kk, Brown:2008sb} using the $1+\log$ slicing and $\Gamma$-driver shift conditions. 
The computational grid set-up is based on the multipatch approach using Llama \cite{PhysRevD.83.044045} and Carpet modules, which enable the mapping and coordinate transformation of multigrid set up from curvilinear coordinates to Cartesian along with adaptive mesh refinement (AMR). It helps to optimally evolve 
the spacetime for a long time in a larger computational domain, and to extract gravitational waves at faraway regions as compared to the Cartesian grid. Individual horizons and common horizons on the numerical grid are found via the method described in \cite{Thornburg:1995cp, Thornburg:2003sf}. We compute the quasi-local quantities on the horizon on an angular grid of size (37, 76) along the longitudinal and latitudinal directions, respectively
The multipole moments in Eq.~\eqref{mass_mult} are obtained from the numerical simulations run using the Einstein Toolkit, using the QuasiLocalMeasures thorn.

We consider non-spinning binary black hole systems with varying mass-ratios 
$q=M_2/M_1$, where $M_{1,2}$ are the masses of the primary and secondary objects. The larger blackhole with mass $M_1$ will be denoted by BH1 and the smaller one with mass $M_2$ by BH2. We will be studying the tidal deformation of 
each of the black holes BH1 and BH2 due to the influence of their companion. 

We use the GW150914 parameter file available from Ref.~\cite{wardell_barry_2016_155394}, as a template for our simulations. For each case, as input parameters, we provide initial separation between the two punctures $D$, mass ratio $q$ and radial and azimuthal linear momentum $p_r$, $p_{\phi}$ respectively, while keeping the total horizon mass $M=M_1+M_2$ of the system to be $1.0$ in units of $c=G=M=1$. Parameters are listed in Table~\ref{tab:ic_qc0}. We compute the corresponding initial locations, the $x$, $y$, $z$ components of linear momentum for both black holes, and grid refinement levels, etc., before generating the initial data and evolving it. We chose non-spinning cases ranging between $q=1.0$ to $0.4$, based on the initial parameters listed in \cite{Healy:2014yta,PhysRevD.95.024037}. Our simulations match very well with the catalog simulations \cite{RITcatalog:web}, having merger time discrepancies of less than a few percent.
\begin{table*}
\begin{tabular}{|p{1.8cm}|p{1.8cm}|p{1.8cm}|p{1.8cm}|p{1.8cm}|p{1.8cm}|}
\hline
\multicolumn{6}{|c|}{Non-spinning BBH Simulations} \\
\hline
\hline
Mass ratio &$d$ &$M_{1}$ &$M_{2}$ &$p_r$ &$p_t$\\
\hline

1.0     & 11.0        & 0.5       & 0.5       & -7.220e-04 & 0.09019  \\
0.85    & 12.0        & 0.54051   & 0.4595    & -5.290e-04 & 0.08448  \\
0.75    & 11.0        & 0.5714    & 0.4286    & -6.860e-04 & 0.08828  \\
0.6667  & 11.75       & 0.6       & 0.4000    & -5.290e-04 & 0.08281  \\
0.50    & 11.0        & 0.6667    & 0.3333    & -5.720e-04 & 0.0802   \\
0.40    & 11.25       & 0.7143    & 0.2857    & -4.500e-04 & 0.07262  \\
\hline
\end{tabular}
\caption{Initial parameters for non-spinning binary black holes with quasi-circular orbits. Here, $q=M_2/M_1$ is the mass-ratio, $d$ is the initial separation between the two holes, $p_r$ and $p_t$ are the radial and tangential (to the orbit) momenta in the initial data, respectively.}
\label{tab:ic_qc0}
\end{table*}

\section{Computing the tidal coefficients}
\label{sec:com_tidal_coeffs}

We now describe a method to compute the leading and sub-leading order tidal coefficients appearing in the expressions for the expansion of the perturbed multipole moments numerically using the aforementioned simulations of BBH mergers. 

To distinguish between the black hole that is tidally deformed and the black hole that sources the tidal field we use $M_{td}$ for the former and $M_{tf}$ for the latter. Thus, $M_{td}$ can be $M_1$ and $M_{tf}$ can be $M_2$ or vice-versa depending on whether the tidal deformation of $M_1$ is being studied or that of $M_2$. We denote an alternate-mass ratio defined using $M_{td}$ and $M_{tf}$ by $\gamma$:
\begin{equation}
  \gamma = \dfrac{M_{tf}}{M_{td}}
\end{equation}
As opposed to the mass ratio defined earlier in Table~\ref{tab:ic_qc0}, $\gamma$ takes values in the range $[0.4, 2.25]$.

\subsection{Identification of the leading order term}

We begin by noting that Eq.~\eqref{exp} is a power-series in the inverse of the distance of separation of the two orbiting black holes. We postulate that there would be no term at order less than $1/d^3$ (i.e., no term involving a lower exponent on $1/d$, such as $1/d^2$ or $1/d$) since the Newtonian tidal force enters at the order of $1/d^3$. Therefore, the leading term is ${\cal O}(1/d^3)$. The sub-leading order term is ${\cal O}(1/d^4)$ and is responsible for post-Newtonian tidal influences. Henceforth, we will use \emph{third-order} and \emph{fourth-order} to refer to the exponent on the distance, with the third-order term being the leading and the fourth-order being the sub-leading term.  

The expression for the perturbative expansion for $n=2$ mass multipole moment at the leading (Model~A), and leading and sub-leading (Model~B) order perturbations due to $M_{tf}$ are, respectively, 
\begin{equation}
  \mathbf{Model \quad A:}\quad \dfrac{\mathcal{M}_2}{M_{td}^3} = \dfrac{a_3 ^{(2)}}{d^3} + {\rm const.}\,
  \label{mass_mult_a3}
\end{equation}
and 
\begin{equation}
  \mathbf{Model \quad B:}\quad \dfrac{\mathcal{M}_2}{M_{td}^3} = \dfrac{a_3 ^{(2)}}{d^3} + \dfrac{a_4^{(2)}}{d^4} + {\rm const.} \label{mass_mult_a34}
\end{equation}
Here, the tidal coefficients are related to $a_3^{(2)}$ and $a_4^{(2)}$ through:
\begin{align}
a_3^{(2)} =& M_{td} M_{tf}^2 \alpha^{(2)} _{12} + M_{td}^2 M_{tf} \alpha^{(2)} _{21}\,, \label{a3_exp}  \\
a_4^{(2)} =& \alpha^{(2)} _{13} M_{td} M_{tf} ^3 + \alpha^{(2)}_{22} M_{td}^2 M_{tf}^2 + \alpha^{(2)}_{31} M_{td}^3 M_{tf} \,.
\label{a34_exp}  
\end{align}
Note, that in the case of an isolated non-spinning black hole, the horizon geometry is spherically symmetric, therefore, the $n=2$ mass multipole moment is zero, resulting in $\delta{\mathcal{M}}_2 = \mathcal{M}_2$. However, in the above model, we include an overall constant in the fits to allow for any systematic errors in the numerical computation of the multipole moment, and to possibly take into account the fact that the initial data is described by punctures and not real black holes. We use the distance measure which is computed by the simple Euclidean separation between the geometric centroids of the two black holes at every time step in the simulation. This agrees well with the physical distance measures between the black holes. More details can be found in Appendix~\ref{sec:appendix}.

We will now detail the procedure for computing the tidal coefficients $\alpha^{(2)} _{12}$, $\alpha^{(2)} _{21}$, $\alpha^{(2)} _{13}$, $\alpha^{(2)}_{22}$ and $\alpha^{(2)}_{31}$ using these models. Thereafter, we compare the fits obtained. The superscript $(2)$ is used to denote the fact that we are analyzing $n=2$ mass multipole moment; since this paper deals exclusively with $n=2$ mass multipolar deformations, it will be dropped henceforth. Resolving higher $n$ multipole moments would require a sufficiently fine grid on the horizon. Due to computational limitations, we restrict our analysis to $n=2$ multipolar deformations in this work.

\subsection{The Fitting procedure}

To estimate the tidal coefficients we follow a two-step fitting procedure:
\begin{enumerate}
  \item Fit of the multipole moment data to distance data, separately to each of the simulations and black holes, and estimating the coefficients $a_3$ and $a_4$ using the model Eq.~\eqref{mass_mult_a34}.
  \item The re-fitting of all values of the best-fit parameters from the above step to the masses $M_{td}$ and $M_{tf}$, in a combined manner, to the models in Eqs.~\eqref{a3_exp} and \eqref{a34_exp} to obtain the tidal coefficients $\alpha_{ij}$.
\end{enumerate}
We carry out linear regression of the numerical data in the two steps to fit the numerical data to the models. We minimize a least-squares objective function in the process. We directly compute a least-squares objective function on a grid in the parameter space and locate the minimum. Since this method is computationally expensive, we follow the two step procedure mentioned above. 

For carrying out the regression, we relate the models and the data in a matrix form as
\begin{equation}
  Y = X A \,,
\end{equation}
where $Y$ is a matrix of dimensions ($K$, $1$) of the left-hand side of the respective model, $X$ is the matrix of data points of dimensions ($K$, $L$) and $A$ the matrix of the parameters of dimensions ($L$, 1). For instance, in the fitting procedure of step one, $K$ is equal to the number of data points in the time-series data of the multipole moment/distance and $L$ is the number of parameters in the model (which is 3 for Model~B). 

We use the following least-squares objective function in the minimization procedure (the summation convention is assumed on repeating indices):
\begin{equation}
  \mathcal{L}(A) = \sum_i (Y_i - X_{ij} A_j)^2 \,,
  \label{lsof}
\end{equation}
Here $X_{ij}A_{j}$ is the prediction from the respective linear models. The linearization of the above models in Eqs.~\eqref{mass_mult_a3}, \eqref{mass_mult_a34}, \eqref{a3_exp}, and \eqref{a34_exp} is done by assuming the fitting parameters as coefficients of linear variables. To exemplify this, let us consider the two steps of the fitting procedure for Model~B. In step one, the linearized model corresponding to Eq.~\eqref{mass_mult_a34} would be written as:
\begin{equation}
  \dfrac{\mathcal{M}_2}{M_{td}^3} = a_3 x_3 + a_4 x_4 + {\rm const.}\,,
\end{equation}
where $x_3 = 1/d^3$ and $x_4=1/d^4$ are the linear fitting variables. For re-fitting, Eqs.~\eqref{a3_exp} and \eqref{a34_exp} would be used. The re-fitting model corresponding to Eq.~\eqref{a34_exp} would be written as:
\begin{equation}
  a_4 = \alpha_{13} \mu_{13} + \alpha_{22} \mu_{22} + \alpha_{31} \mu_{31}\,,
\end{equation}
where $\mu_{13} = M_{td} M_{tf}^3$, and so on.

\subsubsection{Error estimation}

The best-fit parameter values (denoted by $\hat{A}$)
are those that minimize the least-squares objective function Eq.~\eqref{lsof} for the respective models. We compute the errors on the best-fit parameter values from the diagonal components of the variance-covariance matrix $\mathcal{C}$ of the data, and the fit residue $\sigma_{fit}$:
\begin{equation}
  \mathcal{C} = X^T X
\end{equation}
and
\begin{equation}
  \sigma_{fit} = \sqrt{\mathcal{L}(\hat{A})}
\end{equation}
as
\begin{equation}
  \sigma(\hat{A}) = \sigma_{fit} \times Diag(\mathcal{C})\,.
\end{equation}
Here, the matrix of best-fit parameters is denoted by $\hat{A}$ and their standard error of estimates by $\sigma(\hat{A})$. We will refer to the individual elements of these matrices by $\hat{A}_j$ and $\sigma_j$ respectively. Since these parameters can correspond to the fitting procedure of either step one or step two, we will explicitly mention which step we are referring to during their usage.

\section{Results}
\label{sec:results}
\begin{table*}
\begin{tabular}{|p{4.5cm}|p{1.8cm}|p{1.8cm}|p{1.8cm}|p{1.8cm}|p{1.8cm}|}
\hline
\multicolumn{6}{|c|}{Tidal coefficients} \\
\hline
Model                           &   $\alpha_{12}$   &   $\alpha_{21}$    &   $\alpha_{13}$   &   $\alpha_{22}$   &  $\alpha_{31}$   \\
\hline
Third-order model               &   $-0.15 \pm 0.21$  &   $-3.43 \pm 0.21$   &   NA              &   NA              &   NA             \\
Third and fourth-order model    &   $-0.89 \pm 0.21$  &   $-5.45 \pm 0.21$   &   $1.79 \pm 1.72$   &   $5.3 \pm 3.65$    &   $4.68 \pm 1.72$  \\
\hline
\end{tabular}
\caption{Tidal coefficient values estimated from a re-fit of the fit coefficients $a_3$ and $a_3, a_4$ in Eq.~\eqref{mass_mult_a3} and Eq.~\eqref{mass_mult_a34}, respectively.}
\label{tab:fit_coeffs_com}
\end{table*}
\begin{figure*}[htp]
    \centering
    \begin{subfigure}[t]{0.49\textwidth}
    \centering
        \includegraphics[width=\columnwidth]{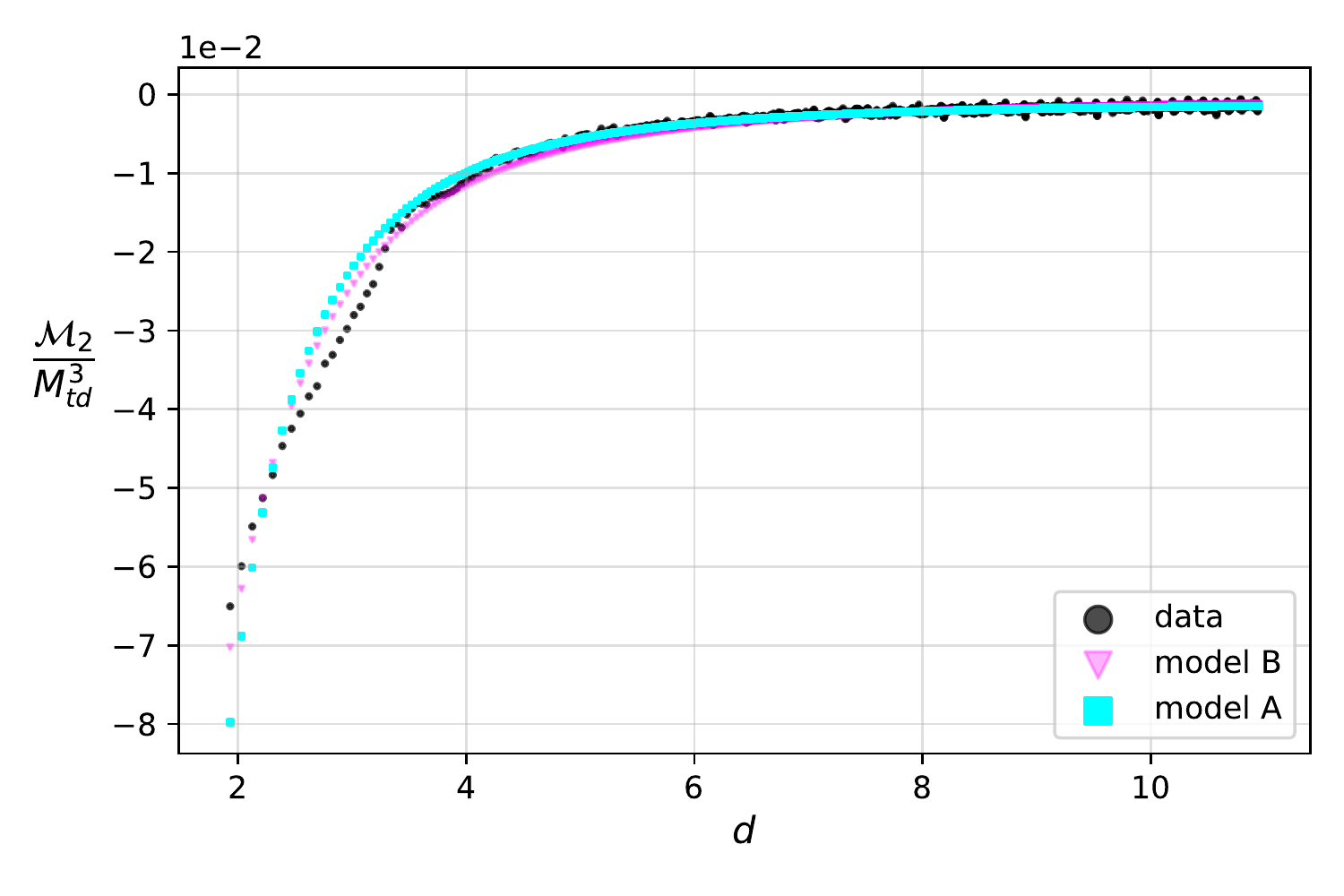}
    \end{subfigure}
    \hfill
    \begin{subfigure}[t]{0.49\textwidth}
        \centering
       \includegraphics[width=\columnwidth]{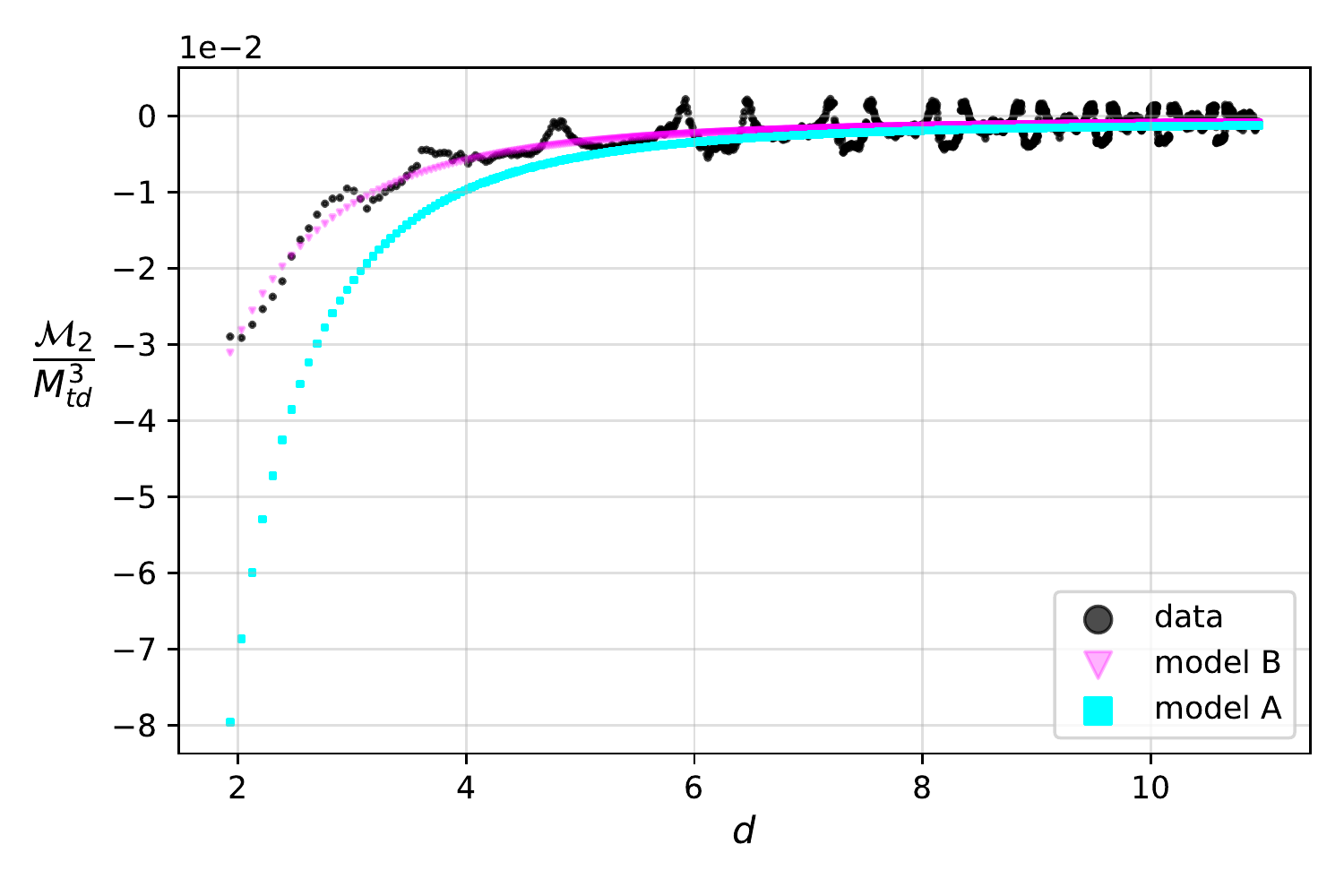}
    \end{subfigure}
    \hfill
     \centering
    \begin{subfigure}[t]{0.49\textwidth}
        \centering
        \includegraphics[width=\columnwidth]{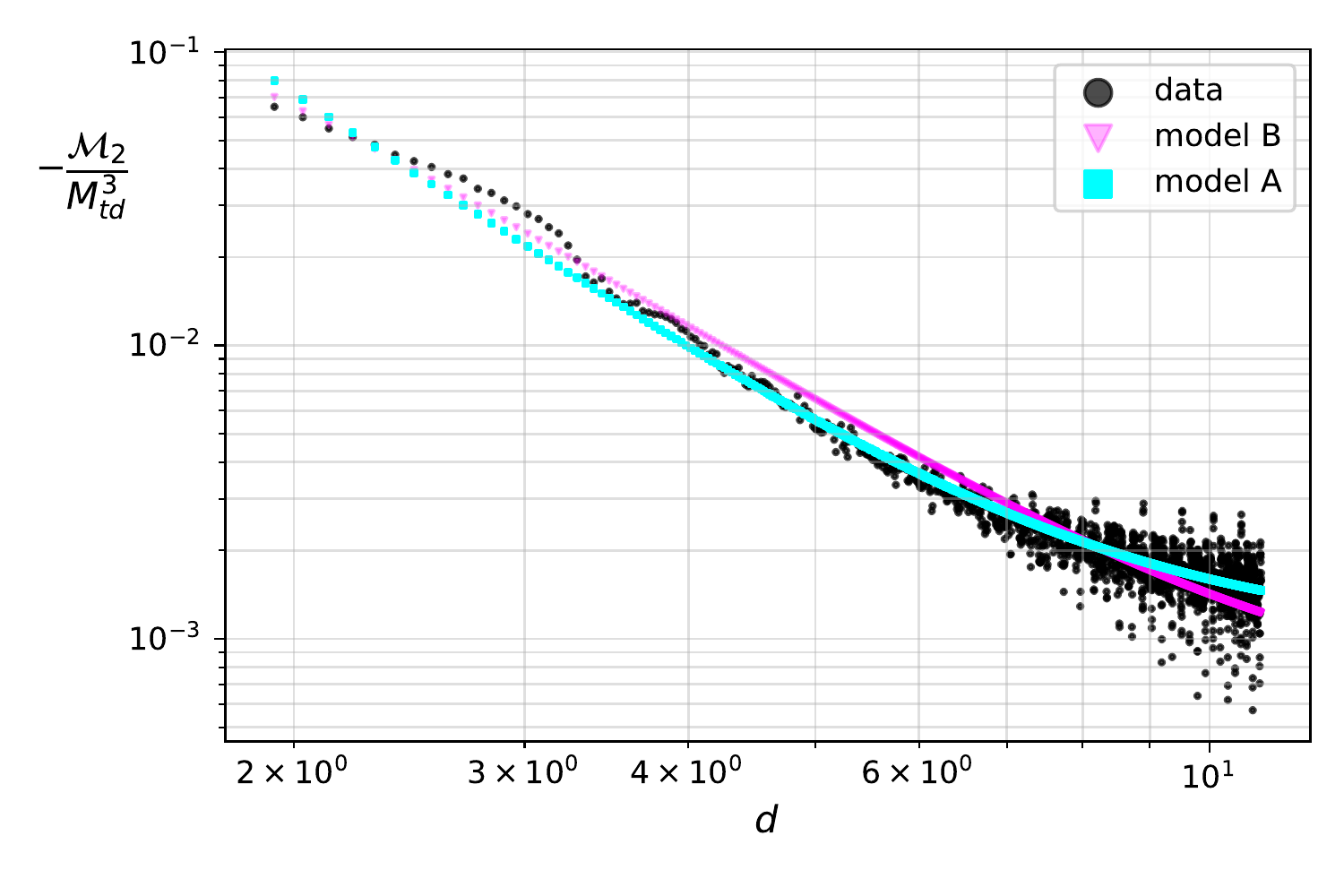}
    \end{subfigure}
    \hfill
   \begin{subfigure}[t]{0.49\textwidth}
       \centering
    \includegraphics[width=\columnwidth]{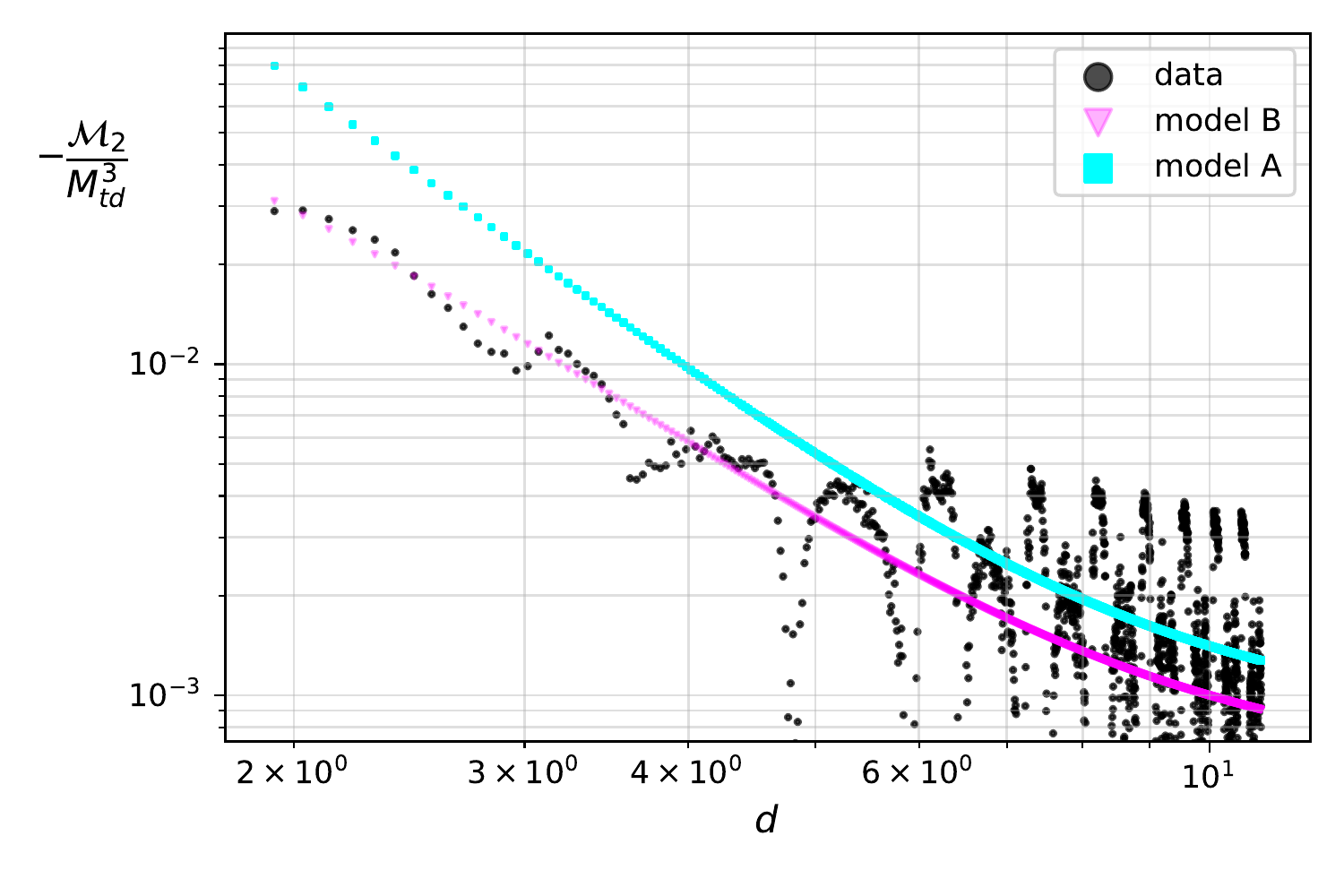}
  \end{subfigure}
\caption{The fit of the $n = 2$ mass multipole moment of BH1 (left) and BH2 (right) for the simulation $q=0.4$. \textbf{Top}: The multipole moment data is plotted against the distance $d$.  
\textbf{Bottom}: The multipole moment data and the distance $d$ are plotted on a logarithmic scale. Note that the variable on the $y$-axis of the bottom figure contains the negative of the multipole moment.
In these two figures, the model where only the leading (third) order term is included (namely, Model~A of Eq.~\eqref{mass_mult_a3}) is shown in cyan (squares) whereas the model where the leading and sub-leading order terms have been included (namely, Model~B of Eq.~\eqref{mass_mult_a34}) is shown in magenta (triangles). The data is in black (dots).}
\label{fig:mlfits_34_q0p4}
\end{figure*}
\begin{figure}[htp]
   \centering
       \includegraphics[width=\columnwidth]{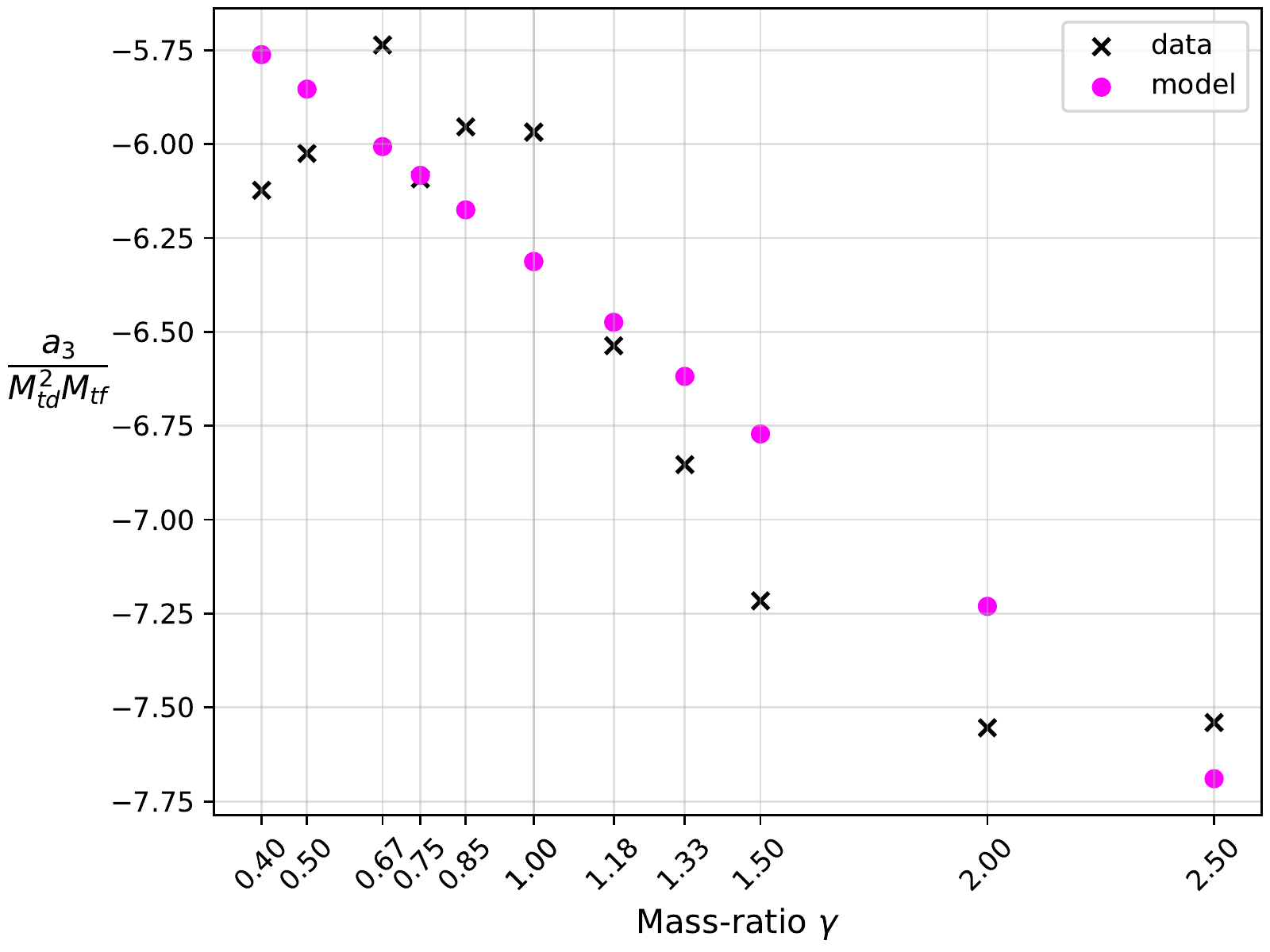}
\caption{The re-fit of the coefficient $a_3$ to obtain the tidal coefficients in Eq.~\eqref{a3_exp} (Model~B). Here 
 $a_3/(M_{td}^2 M_{tf})$ is plotted against $\gamma$. The data points are individual best fit parameters of Model~B in Eq.~\eqref{a34_exp} for both black holes in each simulation. The data points for the fit are in black (crosses) and the best fit model values are in magenta (dots).}
\label{fig:refits_4_3order_12_com}
\end{figure}
\begin{figure}[htp]
       \includegraphics[width=\columnwidth]{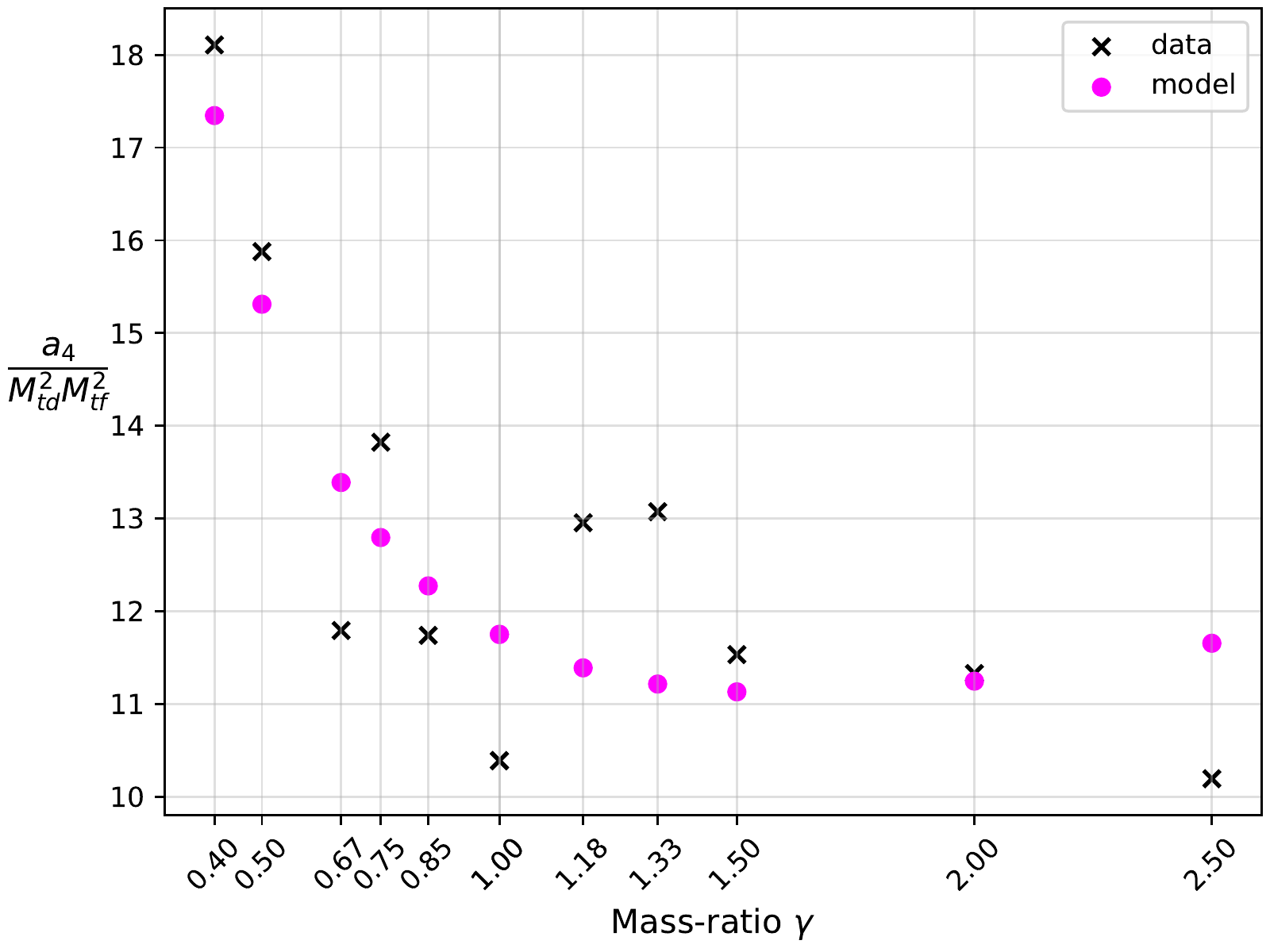}
\caption{The re-fit of $a_4$ to the model of Eq.~\eqref{a34_exp}, with three terms, 
for the simulations $q\geq 0.4$ (Model~B). Here,
$\dfrac{a_4}{M_{td}^2 M_{tf}^2}$ is plotted against $\gamma$. The data points are individual best fit parameters of Model~B in Eq.~\eqref{a34_exp} for both black holes in each simulation. The data points for the fit are in black (crosses) and the best fit model values are in magenta (dots).}
\label{fig:refits_4_4order_123_com}
\end{figure}
We obtain the best-fit parameters for the two models Eq.~\eqref{mass_mult_a3} and Eq.~\eqref{mass_mult_a34} using the two-step procedure mentioned in the preceding section and tabulate the results in Table~\ref{tab:fit_coeffs_com}. In this table, the final results of the step two of the fitting procedure (i.e. re-fits to the models in Eqs.~\eqref{a3_exp} and \eqref{a34_exp}) are denoted in the format $\hat{A}_j \pm \sigma_{j}$: the best-fit parameter value and the respective standard error. 

In Fig.~\ref{fig:mlfits_34_q0p4}, we show the fit of the $n=2$ mass multipole moment to the two models Eq.~\eqref{mass_mult_a3} (Model~A) and Eq.~\eqref{mass_mult_a34} (Model~B) for $q=0.4$ simulation data. In these plots, we show the data points and the best fitting model i.e. the values of the $n=2$ multipole moment $\mathcal{M}_2$ predicted by the models in Eqs.~\eqref{mass_mult_a3} and \eqref{mass_mult_a34} for the best fit parameter values $\hat{A}$. In the top panel, we plot the multipole moment data {\it vs} the distance of separation for the more massive black hole (left) and the less massive black hole (right). In the bottom panel, we plot the same in logarithmic scale on both the axes.

In Figs.~\ref{fig:refits_4_3order_12_com} and ~\ref{fig:refits_4_4order_123_com}, we show the results of step two of the fitting procedure (re-fitting results) of the parameters $a_3$ and $a_4$ of Model~B appearing in Eq.~\eqref{a3_exp} and Eq.~\eqref{a34_exp}, respectively, which were obtained from step one of the fit for the set of binary black hole configurations as listed in Table~\ref{tab:ic_qc0}. In these plots, we show the data points and the best fitting model i.e. the values of $a_3$ and $a_4$ predicted by the models in Eqs.~\eqref{a3_exp} and \eqref{a34_exp} for the best fit parameter values $\hat{A}$.

Some comments are in order regarding these fits. 

First, as can be clearly seen, Model~B, consisting of both $1/d^3$ and $1/d^4$ terms, fits the data better in comparison to  Model~A (which has only the $1/d^3$ term).

Second, although the overall fit to the data of Model~A is worse than that of Model~B, nevertheless the former model fits the data relatively well in the early inspiral phase, i.e., at large distances of separation between the black holes, especially, for the heavier black hole (left panel of Fig.~\ref{fig:mlfits_34_q0p4}).

Since the system spends more time in the early inspiral phase (at large distances of separation), there are more points at large $d$ to which the model fits well. However, this model fails to explain the data as the black holes close in. This is expected because the sub-leading term in Eq.~\eqref{mass_mult_a34} is expected to become increasingly important at low separation distances $d$. This is 
why we expect Model~B to be more accurate.

Third, the quality of fit of the leading order model to the data: the fit of Model~A is visibly worse, especially, for the smaller black hole (right panels of Fig.~\ref{fig:mlfits_34_q0p4}). From a perturbative point of view, we can intuitively explain this as follows. The change in the geometry of the smaller black hole due to the presence of the larger black hole is more than the change in geometry, as quantified by the multipole moments, of the larger black hole due to the smaller one, as the larger black hole is more massive leading to more variations in its multipole moment. 
The sub-leading (fourth) order term contains higher powers of the perturbing black hole's mass $M_{tf}$ than at the leading (third) order ($M_{tf} ^3$ as opposed to $M_{tf} ^ 2$ in Eq.~\eqref{a3_exp}~and~ \eqref{a34_exp}).
Therefore, the effect of the fourth-order term will be easily visible in the multipole moment of the smaller black hole $M_2$ (see right panels in Fig.~\ref{fig:mlfits_34_q0p4}) and the fit between data and Model~A is visibly worse in the plots for the smaller black hole.

Fourth, the fourth-order term has the opposite sign to that of the third. This is visible in the same plots Fig.~\ref{fig:mlfits_34_q0p4} at small values of $d$. The effective slope of the data points deviates from $1/d^3$ behaviour and reduces in magnitude as the distance between the black holes reduces. Therefore, the fourth-order term is responsible for reducing the overall tidal force involving the leading and sub-leading term, consistent with the velocity independent post-Newtonian correction to the Newtonian tidal force (see Ref.~\cite{xupaik2016} and references therein):
\begin{equation}
    \vert\vert \mathbf{F}_{1PN} \vert\vert \sim \dfrac{M_{td}^2 M_{tf}}{d^3}\,.
\end{equation}
One feature of the model in Eq.~\eqref{a34_exp} is that there are three independent terms at the fourth-order in distance whereas there seems to be just one term in the expression for the 1PN correction to the Newtonian gravitational force. It was verified that all three terms were necessary to explain the data points, which have a quadratic dependence on the mass-ratio $q$ -- as can be seen in the bottom panel of Fig.~\ref{fig:refits_4_4order_123_com}, and obtain a good fit to the data. This needs to be probed further in the future.

Fifth, there is room for some error at the fourth-order in $d$. At this 1PN order, terms involving the relative velocities of the black holes would enter into the expansion of Eq.~\eqref{mass_mult_a34}. These terms might be contributing more to the tidal effects as the mass-ratio decreases.

Therefore, in order to analyze the multipole moment data to infer the tidal coefficients at further smaller mass-ratios than $q=0.4$, one may need to take into account new terms involving velocities in the perturbative multipole expansion. 

To further check the precision of these fits, we considered multiple sets of data having various combinations of our simulations, with additional configurations for intermediate $q=\{0.6, 0.7\}$;  excluding a few configurations randomly; or even adding/removing a few lower $q \leq 0.6$. Across these sets, we find $\alpha_{12}$ to be $\approx -1.0$ and $\alpha_{21} \approx -5.3$. Based on the results, we conclude that the tidal Love numbers at the leading order ($\alpha_{12}$ and $\alpha_{21}$) are estimated at a higher precision than for those at the sub-leading order ($\alpha_{13}$, $\alpha_{22}$ and $\alpha_{31}$). This is reasonable since we are attempting to estimate the coefficients of the sub-leading order term, to which there are three contributions (Eq.~\eqref{a34_exp}). The contribution of the fourth-order term is not visible until the onset of the late inspiral phase of the evolution -- i.e., at smaller values of $d$ (as can be seen in Fig.~\ref{fig:mlfits_34_q0p4}). It has the largest impact during this phase, where the number of data points is significantly less than that in the early inspiral phase, thus making it difficult to estimate it at the same precision.

The estimates for the tidal Love numbers at the fourth order may be expected to be computed with more accuracy by increasing the number of simulations, and the resolution of the runs.

\section{Discussion and Conclusions}
\label{sec:conclusions}

The primary goal of this paper was to study the tidal effects on horizons during the inspiral phase of binary black hole mergers. Past studies have examined this problem using the perturbations of the asymptotic gravitational fields, but this is the first time that tidal effects have been explored for a binary system comprising of similar masses using full numerical relativity simulations, and to deduce a black hole's tidal coefficients.

We used source multipoles computed on the marginally outer trapped surfaces  of the space-like dynamical horizons of both the black holes for this analysis. We simulated non-spinning BBH with varying mass ratios $q \in [0.4, 1]$. We defined a convenient set of tidal coefficients to characterize the tidal deformability of black holes in a binary system using the source mass multipole moments of a dynamical horizon. This approach does not involve the field multipole moments that are defined far away from the system, nor do they involve assuming a matter distribution whose compactness limit is taken to describe the black hole case. 
Assumptions regarding the stationarity/ slow evolution of the tides were not made either.  By definition, these numbers are dimensionless, and should be independent of the system considered, i.e., same for all black holes (of any mass and spins) in General Relativity. We computed five tidal Love numbers which characterize the $n=2$ mass multipolar deformations of the dynamical horizon geometry : two at the leading order and three at the sub-leading order.

These results show in explicit detail, as noted in existing literature, that, although the Love numbers that characterize the tidal deformations of the gravitational fields of non-spinning black holes far away from the system are zero, the corresponding Love numbers that characterize the strong field tidal deformation of the horizon geometry in the strong field regime do not vanish.

The relations in Eqs.~\eqref{mass_mult_a34},~ \eqref{a3_exp}~and~\eqref{a34_exp}, together with the best-fit parameters listed in Table~\ref{tab:fit_coeffs_com}, show how the mass multipole moment of the horizon evolves in relation to the dynamics of the system in a binary black hole merger scenario, apart from describing the evolution of tidal deformations. We find that the evolution of the multipole moments can be described quite accurately by the model in Eq.~\eqref{mass_mult_a34} up to the merger. 

Here, it must be noted that in simulations of lower mass-ratios, we notice an increase in the error of our fits to numerical data. This can be attributed to the increasing importance of the terms involving relative velocities of the black holes, and terms involving higher powers of $1/d$. In principle, more tidal Love numbers at successively higher orders can be computed in the manner described here, given sufficient number of numerical simulations are sufficient in number, and are carried out at sufficiently high resolution. It will be important to examine those cases with higher resolution runs in the future.

Analogous to the treatment of neutron star tidal deformability, a source-independent definition of the tidal deformation of black hole horizon using the source multipole moments of the horizon must also be possible. This would involve studying the perturbation of the geometry of an otherwise isolated horizon due to the external perturbing fields, and finding a relation between them -- a quest that may be pursued in a future work. 

\acknowledgments

This research was supported in part by a grant from the Navajbai Ratan Tata Trust. V.P is funded by Shyama Prasad Mukherjee Fellowship, (CSIR).
The numerical simulations for this paper were performed on the Pegasus cluster at The Inter-University Centre for Astronomy and Astrophysics, Pune, India (IUCAA).

\appendix

\section{Distance measure}
\label{sec:appendix}

An estimate of the distance between the two black holes is required to calculate the tidal coefficients. We now discuss various choices for this distance measure and describe how we compute it from the numerical simulations. In the simulations we carried out, the numerical evolution uses a coordinate system described in \cite{Thornburg_2003}. We define and compute the following measures of distances of separation between the black holes:

\begin{itemize}
    \item The Euclidean distance between the geometric centroids of the individual apparent horizons of the two black holes. 
    \item Newtonian proper distance. 
    \item Post-Newtonian (PN) distances up to second-order (PN1, PN2).
\end{itemize}

We explain each of these definitions below.

\subsection{Euclidean distance}
    
We use the coordinate locations of the geometric centroids ($(x_1, y_1, z_1)$ and $(x_2, y_2, z_2)$) of the individual horizons of the two black holes to define a simple Euclidean distance measure.
The distance is then defined by: 
\begin{align}
    d = \sqrt{(x_1 - x_2)^2 + (y_1 - y_2)^2 + (z_1 - z_2)^2} \,.
\end{align}

\subsubsection{Newtonian proper distance}

Kepler's third law for orbiting binaries can be used to calculate the physical distance of separation between the center of masses of the orbiting objects:
\begin{equation}
    T^2 = k d^3 \,.
\end{equation}

We compute the instantaneous time period of the quasi-circular configuration by using the instantaneous frequency of the gravitational waves emitted from the system. We use the extracted gravitational radiation at $100 M$ from the  center of the system for this purpose.

The gravitational waves emitted from the system provided by the WaveExtractCPM thorn in the Llama package in the Einstein toolkit infrastructure. 

\subsubsection{Post-Newtonian distances}
Using a method similar to the one described above for computing the Newtonian proper distance, we can compute proper distances to any Post Newtonian order by using an appropriate PN corrected version of Kepler's third law. Using this method, we compute PN1 and PN2 distances between the centers of the black holes. 

\subsubsection{Choosing the distance measure}

\begin{figure}[htp]
   \centering
    \includegraphics[width=\columnwidth]{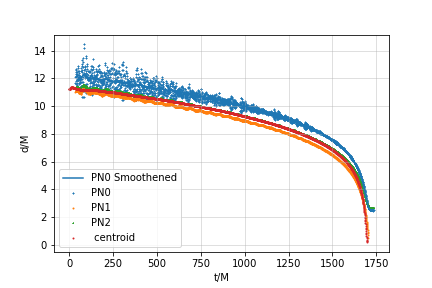}
\caption{The various distance measures for the $q=0.4$ simulation. }
\label{fig:dist_measures_q0p4}
\end{figure}

The plot for these distance measures for one of our simulations is shown in Fig.~\ref{fig:dist_measures_q0p4} for illustration. 

It is observed that although the simple Euclidean distance measure seems less appropriate in comparison with the rest of the proper distance measures, it is the measure closest to the second-order Post-Newtonian distance measure. For any given simulation, the maximum cumulative (RMS) deviation between the simple Euclidean and other distance measures was found to be less than few times the initial separation. Further, the cumulative deviation with the PN2 distance measure was found to be the least for the simple Euclidean distance measure. Therefore, throughout the analysis, we choose to work with this distance measure. 

The simple Euclidean distance measure also closely matches the Newtonian evolution of the separation between the two black holes. This can be verified by means of simple fitting (see also \cite{etk_student_guide_2020}).
    
\bibliography{references.bib}

\begin{thebibliography}{56}%
\makeatletter
\providecommand \@ifxundefined [1]{%
 \@ifx{#1\undefined}
}%
\providecommand \@ifnum [1]{%
 \ifnum #1\expandafter \@firstoftwo
 \else \expandafter \@secondoftwo
 \fi
}%
\providecommand \@ifx [1]{%
 \ifx #1\expandafter \@firstoftwo
 \else \expandafter \@secondoftwo
 \fi
}%
\providecommand \natexlab [1]{#1}%
\providecommand \enquote  [1]{``#1''}%
\providecommand \bibnamefont  [1]{#1}%
\providecommand \bibfnamefont [1]{#1}%
\providecommand \citenamefont [1]{#1}%
\providecommand \href@noop [0]{\@secondoftwo}%
\providecommand \href [0]{\begingroup \@sanitize@url \@href}%
\providecommand \@href[1]{\@@startlink{#1}\@@href}%
\providecommand \@@href[1]{\endgroup#1\@@endlink}%
\providecommand \@sanitize@url [0]{\catcode `\\12\catcode `\$12\catcode
  `\&12\catcode `\#12\catcode `\^12\catcode `\_12\catcode `\%12\relax}%
\providecommand \@@startlink[1]{}%
\providecommand \@@endlink[0]{}%
\providecommand \url  [0]{\begingroup\@sanitize@url \@url }%
\providecommand \@url [1]{\endgroup\@href {#1}{\urlprefix }}%
\providecommand \urlprefix  [0]{URL }%
\providecommand \Eprint [0]{\href }%
\providecommand \doibase [0]{http://dx.doi.org/}%
\providecommand \selectlanguage [0]{\@gobble}%
\providecommand \bibinfo  [0]{\@secondoftwo}%
\providecommand \bibfield  [0]{\@secondoftwo}%
\providecommand \translation [1]{[#1]}%
\providecommand \BibitemOpen [0]{}%
\providecommand \bibitemStop [0]{}%
\providecommand \bibitemNoStop [0]{.\EOS\space}%
\providecommand \EOS [0]{\spacefactor3000\relax}%
\providecommand \BibitemShut  [1]{\csname bibitem#1\endcsname}%
\let\auto@bib@innerbib\@empty
\bibitem [{\citenamefont {Binnington}\ and\ \citenamefont
  {Poisson}(2009)}]{binningtonpoisson2009}%
  \BibitemOpen
  \bibfield  {author} {\bibinfo {author} {\bibfnamefont {T.}~\bibnamefont
  {Binnington}}\ and\ \bibinfo {author} {\bibfnamefont {E.}~\bibnamefont
  {Poisson}},\ }\href {\doibase 10.1103/PhysRevD.80.084018} {\bibfield
  {journal} {\bibinfo  {journal} {Phys. Rev. D}\ }\textbf {\bibinfo {volume}
  {80}},\ \bibinfo {pages} {084018} (\bibinfo {year} {2009})}\BibitemShut
  {NoStop}%
\bibitem [{\citenamefont {{Brooker}}\ and\ \citenamefont
  {{Olle}}(1955)}]{1955MNRAS.115..101B}%
  \BibitemOpen
  \bibfield  {author} {\bibinfo {author} {\bibfnamefont {R.~A.}\ \bibnamefont
  {{Brooker}}}\ and\ \bibinfo {author} {\bibfnamefont {T.~W.}\ \bibnamefont
  {{Olle}}},\ }\href {\doibase 10.1093/mnras/115.1.101} {\bibfield  {journal}
  {\bibinfo  {journal} {MNRAS}\ }\textbf {\bibinfo {volume} {115}},\ \bibinfo
  {pages} {101} (\bibinfo {year} {1955})}\BibitemShut {NoStop}%
\bibitem [{\citenamefont {Flanagan}\ and\ \citenamefont
  {Hinderer}(2008)}]{Flanagan:2007ix}%
  \BibitemOpen
  \bibfield  {author} {\bibinfo {author} {\bibfnamefont {E.~E.}\ \bibnamefont
  {Flanagan}}\ and\ \bibinfo {author} {\bibfnamefont {T.}~\bibnamefont
  {Hinderer}},\ }\href {\doibase 10.1103/PhysRevD.77.021502} {\bibfield
  {journal} {\bibinfo  {journal} {Phys. Rev. D}\ }\textbf {\bibinfo {volume}
  {77}},\ \bibinfo {pages} {021502} (\bibinfo {year} {2008})},\ \Eprint
  {http://arxiv.org/abs/0709.1915} {arXiv:0709.1915 [astro-ph]} \BibitemShut
  {NoStop}%
\bibitem [{\citenamefont {Damour}\ and\ \citenamefont
  {Nagar}(2009{\natexlab{a}})}]{Damour:2009vw}%
  \BibitemOpen
  \bibfield  {author} {\bibinfo {author} {\bibfnamefont {T.}~\bibnamefont
  {Damour}}\ and\ \bibinfo {author} {\bibfnamefont {A.}~\bibnamefont {Nagar}},\
  }\href {\doibase 10.1103/PhysRevD.80.084035} {\bibfield  {journal} {\bibinfo
  {journal} {Phys. Rev. D}\ }\textbf {\bibinfo {volume} {80}},\ \bibinfo
  {pages} {084035} (\bibinfo {year} {2009}{\natexlab{a}})},\ \Eprint
  {http://arxiv.org/abs/0906.0096} {arXiv:0906.0096 [gr-qc]} \BibitemShut
  {NoStop}%
\bibitem [{\citenamefont {Lattimer}\ and\ \citenamefont
  {Prakash}(2016)}]{Lattimer:2015nhk}%
  \BibitemOpen
  \bibfield  {author} {\bibinfo {author} {\bibfnamefont {J.~M.}\ \bibnamefont
  {Lattimer}}\ and\ \bibinfo {author} {\bibfnamefont {M.}~\bibnamefont
  {Prakash}},\ }\href {\doibase 10.1016/j.physrep.2015.12.005} {\bibfield
  {journal} {\bibinfo  {journal} {Phys. Rept.}\ }\textbf {\bibinfo {volume}
  {621}},\ \bibinfo {pages} {127} (\bibinfo {year} {2016})},\ \Eprint
  {http://arxiv.org/abs/1512.07820} {arXiv:1512.07820 [astro-ph.SR]}
  \BibitemShut {NoStop}%
\bibitem [{\citenamefont {Abbott}\ \emph {et~al.}(2018)\citenamefont {Abbott}
  \emph {et~al.}}]{Abbott:2018exr}%
  \BibitemOpen
  \bibfield  {author} {\bibinfo {author} {\bibfnamefont {B.~P.}\ \bibnamefont
  {Abbott}} \emph {et~al.} (\bibinfo {collaboration} {LIGO Scientific,
  Virgo}),\ }\href {\doibase 10.1103/PhysRevLett.121.161101} {\bibfield
  {journal} {\bibinfo  {journal} {Phys. Rev. Lett.}\ }\textbf {\bibinfo
  {volume} {121}},\ \bibinfo {pages} {161101} (\bibinfo {year} {2018})},\
  \Eprint {http://arxiv.org/abs/1805.11581} {arXiv:1805.11581 [gr-qc]}
  \BibitemShut {NoStop}%
\bibitem [{\citenamefont {De}\ \emph {et~al.}(2018)\citenamefont {De},
  \citenamefont {Finstad}, \citenamefont {Lattimer}, \citenamefont {Brown},
  \citenamefont {Berger},\ and\ \citenamefont {Biwer}}]{De:2018uhw}%
  \BibitemOpen
  \bibfield  {author} {\bibinfo {author} {\bibfnamefont {S.}~\bibnamefont
  {De}}, \bibinfo {author} {\bibfnamefont {D.}~\bibnamefont {Finstad}},
  \bibinfo {author} {\bibfnamefont {J.~M.}\ \bibnamefont {Lattimer}}, \bibinfo
  {author} {\bibfnamefont {D.~A.}\ \bibnamefont {Brown}}, \bibinfo {author}
  {\bibfnamefont {E.}~\bibnamefont {Berger}}, \ and\ \bibinfo {author}
  {\bibfnamefont {C.~M.}\ \bibnamefont {Biwer}},\ }\href {\doibase
  10.1103/PhysRevLett.121.091102} {\bibfield  {journal} {\bibinfo  {journal}
  {Phys. Rev. Lett.}\ }\textbf {\bibinfo {volume} {121}},\ \bibinfo {pages}
  {091102} (\bibinfo {year} {2018})},\ \bibinfo {note} {[Erratum:
  Phys.Rev.Lett. 121, 259902 (2018)]},\ \Eprint
  {http://arxiv.org/abs/1804.08583} {arXiv:1804.08583 [astro-ph.HE]}
  \BibitemShut {NoStop}%
\bibitem [{\citenamefont {Capano}\ \emph {et~al.}(2020)\citenamefont {Capano},
  \citenamefont {Tews}, \citenamefont {Brown}, \citenamefont {Margalit},
  \citenamefont {De}, \citenamefont {Kumar}, \citenamefont {Brown},
  \citenamefont {Krishnan},\ and\ \citenamefont {Reddy}}]{Capano:2019eae}%
  \BibitemOpen
  \bibfield  {author} {\bibinfo {author} {\bibfnamefont {C.~D.}\ \bibnamefont
  {Capano}}, \bibinfo {author} {\bibfnamefont {I.}~\bibnamefont {Tews}},
  \bibinfo {author} {\bibfnamefont {S.~M.}\ \bibnamefont {Brown}}, \bibinfo
  {author} {\bibfnamefont {B.}~\bibnamefont {Margalit}}, \bibinfo {author}
  {\bibfnamefont {S.}~\bibnamefont {De}}, \bibinfo {author} {\bibfnamefont
  {S.}~\bibnamefont {Kumar}}, \bibinfo {author} {\bibfnamefont {D.~A.}\
  \bibnamefont {Brown}}, \bibinfo {author} {\bibfnamefont {B.}~\bibnamefont
  {Krishnan}}, \ and\ \bibinfo {author} {\bibfnamefont {S.}~\bibnamefont
  {Reddy}},\ }\href {\doibase 10.1038/s41550-020-1014-6} {\bibfield  {journal}
  {\bibinfo  {journal} {Nature Astron.}\ }\textbf {\bibinfo {volume} {4}},\
  \bibinfo {pages} {625} (\bibinfo {year} {2020})},\ \Eprint
  {http://arxiv.org/abs/1908.10352} {arXiv:1908.10352 [astro-ph.HE]}
  \BibitemShut {NoStop}%
\bibitem [{\citenamefont {Landry}\ and\ \citenamefont
  {Poisson}(2014)}]{Landry2014}%
  \BibitemOpen
  \bibfield  {author} {\bibinfo {author} {\bibfnamefont {P.}~\bibnamefont
  {Landry}}\ and\ \bibinfo {author} {\bibfnamefont {E.}~\bibnamefont
  {Poisson}},\ }\href {\doibase 10.1103/PhysRevD.89.124011} {\bibfield
  {journal} {\bibinfo  {journal} {Phys. Rev. D}\ }\textbf {\bibinfo {volume}
  {89}},\ \bibinfo {pages} {124011} (\bibinfo {year} {2014})}\BibitemShut
  {NoStop}%
\bibitem [{\citenamefont {Damour}\ and\ \citenamefont
  {Lecian}(2009)}]{Damour:2009va}%
  \BibitemOpen
  \bibfield  {author} {\bibinfo {author} {\bibfnamefont {T.}~\bibnamefont
  {Damour}}\ and\ \bibinfo {author} {\bibfnamefont {O.~M.}\ \bibnamefont
  {Lecian}},\ }\href {\doibase 10.1103/PhysRevD.80.044017} {\bibfield
  {journal} {\bibinfo  {journal} {Phys.Rev.}\ }\textbf {\bibinfo {volume}
  {D80}},\ \bibinfo {pages} {044017} (\bibinfo {year} {2009})},\ \Eprint
  {http://arxiv.org/abs/0906.3003} {arXiv:0906.3003 [gr-qc]} \BibitemShut
  {NoStop}%
\bibitem [{\citenamefont {Pani}\ \emph {et~al.}(2015)\citenamefont {Pani},
  \citenamefont {Gualtieri}, \citenamefont {Maselli},\ and\ \citenamefont
  {Ferrari}}]{Pani:2015hfa}%
  \BibitemOpen
  \bibfield  {author} {\bibinfo {author} {\bibfnamefont {P.}~\bibnamefont
  {Pani}}, \bibinfo {author} {\bibfnamefont {L.}~\bibnamefont {Gualtieri}},
  \bibinfo {author} {\bibfnamefont {A.}~\bibnamefont {Maselli}}, \ and\
  \bibinfo {author} {\bibfnamefont {V.}~\bibnamefont {Ferrari}},\ }\href
  {\doibase 10.1103/PhysRevD.92.024010} {\bibfield  {journal} {\bibinfo
  {journal} {Phys. Rev. D}\ }\textbf {\bibinfo {volume} {92}},\ \bibinfo
  {pages} {024010} (\bibinfo {year} {2015})},\ \Eprint
  {http://arxiv.org/abs/1503.07365} {arXiv:1503.07365 [gr-qc]} \BibitemShut
  {NoStop}%
\bibitem [{\citenamefont {Le~Tiec}\ \emph {et~al.}(2021)\citenamefont
  {Le~Tiec}, \citenamefont {Casals},\ and\ \citenamefont
  {Franzin}}]{LeTiec:2020bos}%
  \BibitemOpen
  \bibfield  {author} {\bibinfo {author} {\bibfnamefont {A.}~\bibnamefont
  {Le~Tiec}}, \bibinfo {author} {\bibfnamefont {M.}~\bibnamefont {Casals}}, \
  and\ \bibinfo {author} {\bibfnamefont {E.}~\bibnamefont {Franzin}},\ }\href
  {\doibase 10.1103/PhysRevD.103.084021} {\bibfield  {journal} {\bibinfo
  {journal} {Phys. Rev. D}\ }\textbf {\bibinfo {volume} {103}},\ \bibinfo
  {pages} {084021} (\bibinfo {year} {2021})}\BibitemShut {NoStop}%
\bibitem [{\citenamefont {Le~Tiec}\ and\ \citenamefont
  {Casals}(2021)}]{LeTiec:2020spy}%
  \BibitemOpen
  \bibfield  {author} {\bibinfo {author} {\bibfnamefont {A.}~\bibnamefont
  {Le~Tiec}}\ and\ \bibinfo {author} {\bibfnamefont {M.}~\bibnamefont
  {Casals}},\ }\href {\doibase 10.1103/PhysRevLett.126.131102} {\bibfield
  {journal} {\bibinfo  {journal} {Phys. Rev. Lett.}\ }\textbf {\bibinfo
  {volume} {126}},\ \bibinfo {pages} {131102} (\bibinfo {year}
  {2021})}\BibitemShut {NoStop}%
\bibitem [{\citenamefont {Damour}\ and\ \citenamefont
  {Nagar}(2009{\natexlab{b}})}]{DamourNagar2009}%
  \BibitemOpen
  \bibfield  {author} {\bibinfo {author} {\bibfnamefont {T.}~\bibnamefont
  {Damour}}\ and\ \bibinfo {author} {\bibfnamefont {A.}~\bibnamefont {Nagar}},\
  }\href {\doibase 10.1103/PhysRevD.80.084035} {\bibfield  {journal} {\bibinfo
  {journal} {Phys. Rev. D}\ }\textbf {\bibinfo {volume} {80}},\ \bibinfo
  {pages} {084035} (\bibinfo {year} {2009}{\natexlab{b}})}\BibitemShut
  {NoStop}%
\bibitem [{\citenamefont {Ashtekar}\ and\ \citenamefont
  {Krishnan}(2004)}]{Ashtekar:2004cn}%
  \BibitemOpen
  \bibfield  {author} {\bibinfo {author} {\bibfnamefont {A.}~\bibnamefont
  {Ashtekar}}\ and\ \bibinfo {author} {\bibfnamefont {B.}~\bibnamefont
  {Krishnan}},\ }\href@noop {} {\bibfield  {journal} {\bibinfo  {journal}
  {Living Rev. Rel.}\ }\textbf {\bibinfo {volume} {7}},\ \bibinfo {pages} {10}
  (\bibinfo {year} {2004})},\ \Eprint {http://arxiv.org/abs/gr-qc/0407042}
  {arXiv:gr-qc/0407042} \BibitemShut {NoStop}%
\bibitem [{\citenamefont {Booth}(2005)}]{Booth:2005qc}%
  \BibitemOpen
  \bibfield  {author} {\bibinfo {author} {\bibfnamefont {I.}~\bibnamefont
  {Booth}},\ }\href {\doibase 10.1139/p05-063} {\bibfield  {journal} {\bibinfo
  {journal} {Can. J. Phys.}\ }\textbf {\bibinfo {volume} {83}},\ \bibinfo
  {pages} {1073} (\bibinfo {year} {2005})},\ \Eprint
  {http://arxiv.org/abs/gr-qc/0508107} {arXiv:gr-qc/0508107} \BibitemShut
  {NoStop}%
\bibitem [{\citenamefont {Hayward}(2000)}]{Hayward:2000ca}%
  \BibitemOpen
  \bibfield  {author} {\bibinfo {author} {\bibfnamefont {S.~A.}\ \bibnamefont
  {Hayward}},\ }in\ \href@noop {} {\emph {\bibinfo {booktitle} {{Recent
  developments in theoretical and experimental general relativity, gravitation
  and relativistic field theories. Proceedings, 9th Marcel Grossmann Meeting,
  MG'9, Rome, Italy, July 2-8, 2000. Pts. A-C}}}}\ (\bibinfo {year} {2000})\
  pp.\ \bibinfo {pages} {568--580},\ \Eprint
  {http://arxiv.org/abs/gr-qc/0008071} {arXiv:gr-qc/0008071 [gr-qc]}
  \BibitemShut {NoStop}%
\bibitem [{\citenamefont {Gourgoulhon}\ and\ \citenamefont
  {Jaramillo}(2006)}]{Gourgoulhon:2005ng}%
  \BibitemOpen
  \bibfield  {author} {\bibinfo {author} {\bibfnamefont {E.}~\bibnamefont
  {Gourgoulhon}}\ and\ \bibinfo {author} {\bibfnamefont {J.~L.}\ \bibnamefont
  {Jaramillo}},\ }\href {\doibase 10.1016/j.physrep.2005.10.005} {\bibfield
  {journal} {\bibinfo  {journal} {Phys. Rept.}\ }\textbf {\bibinfo {volume}
  {423}},\ \bibinfo {pages} {159} (\bibinfo {year} {2006})},\ \Eprint
  {http://arxiv.org/abs/gr-qc/0503113} {arXiv:gr-qc/0503113} \BibitemShut
  {NoStop}%
\bibitem [{\citenamefont {Visser}(2008)}]{Visser:2009xp}%
  \BibitemOpen
  \bibfield  {author} {\bibinfo {author} {\bibfnamefont {M.}~\bibnamefont
  {Visser}},\ }\href@noop {} {\bibfield  {journal} {\bibinfo  {journal} {PoS}\
  }\textbf {\bibinfo {volume} {BHSGRANDSTRINGS2008}},\ \bibinfo {pages} {001}
  (\bibinfo {year} {2008})},\ \Eprint {http://arxiv.org/abs/0901.4365}
  {arXiv:0901.4365 [gr-qc]} \BibitemShut {NoStop}%
\bibitem [{\citenamefont {Jaramillo}(2011)}]{Jaramillo:2011zw}%
  \BibitemOpen
  \bibfield  {author} {\bibinfo {author} {\bibfnamefont {J.~L.}\ \bibnamefont
  {Jaramillo}},\ }\href {\doibase 10.1142/S0218271811020366} {\bibfield
  {journal} {\bibinfo  {journal} {Int. J. Mod. Phys.}\ }\textbf {\bibinfo
  {volume} {D20}},\ \bibinfo {pages} {2169} (\bibinfo {year} {2011})},\ \Eprint
  {http://arxiv.org/abs/1108.2408} {arXiv:1108.2408 [gr-qc]} \BibitemShut
  {NoStop}%
\bibitem [{\citenamefont {Faraoni}\ and\ \citenamefont
  {Prain}(2015)}]{Faraoni:2015pmn}%
  \BibitemOpen
  \bibfield  {author} {\bibinfo {author} {\bibfnamefont {V.}~\bibnamefont
  {Faraoni}}\ and\ \bibinfo {author} {\bibfnamefont {A.}~\bibnamefont
  {Prain}},\ }\href@noop {} {\bibfield  {journal} {\bibinfo  {journal} {Lecture
  Notes in Physics}\ }\textbf {\bibinfo {volume} {907}},\ \bibinfo {pages} {1}
  (\bibinfo {year} {2015})},\ \Eprint {http://arxiv.org/abs/1511.07775}
  {arXiv:1511.07775 [gr-qc]} \BibitemShut {NoStop}%
\bibitem [{\citenamefont {Penrose}(1965)}]{Penrose:1964wq}%
  \BibitemOpen
  \bibfield  {author} {\bibinfo {author} {\bibfnamefont {R.}~\bibnamefont
  {Penrose}},\ }\href {\doibase 10.1103/PhysRevLett.14.57} {\bibfield
  {journal} {\bibinfo  {journal} {Phys. Rev. Lett.}\ }\textbf {\bibinfo
  {volume} {14}},\ \bibinfo {pages} {57} (\bibinfo {year} {1965})}\BibitemShut
  {NoStop}%
\bibitem [{\citenamefont {Pook-Kolb}\ \emph
  {et~al.}(2020{\natexlab{a}})\citenamefont {Pook-Kolb}, \citenamefont
  {Birnholtz}, \citenamefont {Jaramillo}, \citenamefont {Krishnan},\ and\
  \citenamefont {Schnetter}}]{Pook-Kolb:2020zhm}%
  \BibitemOpen
  \bibfield  {author} {\bibinfo {author} {\bibfnamefont {D.}~\bibnamefont
  {Pook-Kolb}}, \bibinfo {author} {\bibfnamefont {O.}~\bibnamefont
  {Birnholtz}}, \bibinfo {author} {\bibfnamefont {J.~L.}\ \bibnamefont
  {Jaramillo}}, \bibinfo {author} {\bibfnamefont {B.}~\bibnamefont {Krishnan}},
  \ and\ \bibinfo {author} {\bibfnamefont {E.}~\bibnamefont {Schnetter}},\
  }\href@noop {} {\  (\bibinfo {year} {2020}{\natexlab{a}})},\ \Eprint
  {http://arxiv.org/abs/2006.03939} {arXiv:2006.03939 [gr-qc]} \BibitemShut
  {NoStop}%
\bibitem [{\citenamefont {Pook-Kolb}\ \emph
  {et~al.}(2020{\natexlab{b}})\citenamefont {Pook-Kolb}, \citenamefont
  {Birnholtz}, \citenamefont {Jaramillo}, \citenamefont {Krishnan},\ and\
  \citenamefont {Schnetter}}]{Pook-Kolb:2020jlr}%
  \BibitemOpen
  \bibfield  {author} {\bibinfo {author} {\bibfnamefont {D.}~\bibnamefont
  {Pook-Kolb}}, \bibinfo {author} {\bibfnamefont {O.}~\bibnamefont
  {Birnholtz}}, \bibinfo {author} {\bibfnamefont {J.~L.}\ \bibnamefont
  {Jaramillo}}, \bibinfo {author} {\bibfnamefont {B.}~\bibnamefont {Krishnan}},
  \ and\ \bibinfo {author} {\bibfnamefont {E.}~\bibnamefont {Schnetter}},\
  }\href@noop {} {\  (\bibinfo {year} {2020}{\natexlab{b}})},\ \Eprint
  {http://arxiv.org/abs/2006.03940} {arXiv:2006.03940 [gr-qc]} \BibitemShut
  {NoStop}%
\bibitem [{\citenamefont {Andersson}\ \emph {et~al.}(2009)\citenamefont
  {Andersson}, \citenamefont {Mars}, \citenamefont {Metzger},\ and\
  \citenamefont {Simon}}]{Andersson:2008up}%
  \BibitemOpen
  \bibfield  {author} {\bibinfo {author} {\bibfnamefont {L.}~\bibnamefont
  {Andersson}}, \bibinfo {author} {\bibfnamefont {M.}~\bibnamefont {Mars}},
  \bibinfo {author} {\bibfnamefont {J.}~\bibnamefont {Metzger}}, \ and\
  \bibinfo {author} {\bibfnamefont {W.}~\bibnamefont {Simon}},\ }\href
  {\doibase 10.1088/0264-9381/26/8/085018} {\bibfield  {journal} {\bibinfo
  {journal} {Class.Quant.Grav.}\ }\textbf {\bibinfo {volume} {26}},\ \bibinfo
  {pages} {085018} (\bibinfo {year} {2009})},\ \Eprint
  {http://arxiv.org/abs/0811.4721} {arXiv:0811.4721 [gr-qc]} \BibitemShut
  {NoStop}%
\bibitem [{\citenamefont {Andersson}\ \emph {et~al.}(2008)\citenamefont
  {Andersson}, \citenamefont {Mars},\ and\ \citenamefont
  {Simon}}]{Andersson:2007fh}%
  \BibitemOpen
  \bibfield  {author} {\bibinfo {author} {\bibfnamefont {L.}~\bibnamefont
  {Andersson}}, \bibinfo {author} {\bibfnamefont {M.}~\bibnamefont {Mars}}, \
  and\ \bibinfo {author} {\bibfnamefont {W.}~\bibnamefont {Simon}},\
  }\href@noop {} {\bibfield  {journal} {\bibinfo  {journal}
  {Adv.Theor.Math.Phys.}\ }\textbf {\bibinfo {volume} {12}} (\bibinfo {year}
  {2008})},\ \Eprint {http://arxiv.org/abs/0704.2889} {arXiv:0704.2889 [gr-qc]}
  \BibitemShut {NoStop}%
\bibitem [{\citenamefont {Pook-Kolb}\ \emph
  {et~al.}(2019{\natexlab{a}})\citenamefont {Pook-Kolb}, \citenamefont
  {Birnholtz}, \citenamefont {Krishnan},\ and\ \citenamefont
  {Schnetter}}]{PhysRevD.100.084044}%
  \BibitemOpen
  \bibfield  {author} {\bibinfo {author} {\bibfnamefont {D.}~\bibnamefont
  {Pook-Kolb}}, \bibinfo {author} {\bibfnamefont {O.}~\bibnamefont
  {Birnholtz}}, \bibinfo {author} {\bibfnamefont {B.}~\bibnamefont {Krishnan}},
  \ and\ \bibinfo {author} {\bibfnamefont {E.}~\bibnamefont {Schnetter}},\
  }\href {\doibase 10.1103/PhysRevD.100.084044} {\bibfield  {journal} {\bibinfo
   {journal} {Phys. Rev. D}\ }\textbf {\bibinfo {volume} {100}},\ \bibinfo
  {pages} {084044} (\bibinfo {year} {2019}{\natexlab{a}})}\BibitemShut
  {NoStop}%
\bibitem [{\citenamefont {Pook-Kolb}\ \emph
  {et~al.}(2019{\natexlab{b}})\citenamefont {Pook-Kolb}, \citenamefont
  {Birnholtz}, \citenamefont {Krishnan},\ and\ \citenamefont
  {Schnetter}}]{PhysRevLett.123.171102}%
  \BibitemOpen
  \bibfield  {author} {\bibinfo {author} {\bibfnamefont {D.}~\bibnamefont
  {Pook-Kolb}}, \bibinfo {author} {\bibfnamefont {O.}~\bibnamefont
  {Birnholtz}}, \bibinfo {author} {\bibfnamefont {B.}~\bibnamefont {Krishnan}},
  \ and\ \bibinfo {author} {\bibfnamefont {E.}~\bibnamefont {Schnetter}},\
  }\href {\doibase 10.1103/PhysRevLett.123.171102} {\bibfield  {journal}
  {\bibinfo  {journal} {Phys. Rev. Lett.}\ }\textbf {\bibinfo {volume} {123}},\
  \bibinfo {pages} {171102} (\bibinfo {year} {2019}{\natexlab{b}})}\BibitemShut
  {NoStop}%
\bibitem [{\citenamefont {Pook-Kolb}\ \emph {et~al.}(2021)\citenamefont
  {Pook-Kolb}, \citenamefont {Hennigar},\ and\ \citenamefont
  {Booth}}]{Pook-Kolb:2021gsh}%
  \BibitemOpen
  \bibfield  {author} {\bibinfo {author} {\bibfnamefont {D.}~\bibnamefont
  {Pook-Kolb}}, \bibinfo {author} {\bibfnamefont {R.~A.}\ \bibnamefont
  {Hennigar}}, \ and\ \bibinfo {author} {\bibfnamefont {I.}~\bibnamefont
  {Booth}},\ }\href@noop {} {\  (\bibinfo {year} {2021})},\ \Eprint
  {http://arxiv.org/abs/2104.10265} {arXiv:2104.10265 [gr-qc]} \BibitemShut
  {NoStop}%
\bibitem [{\citenamefont {Booth}\ \emph {et~al.}(2021)\citenamefont {Booth},
  \citenamefont {Hennigar},\ and\ \citenamefont {Pook-Kolb}}]{Booth:2021sow}%
  \BibitemOpen
  \bibfield  {author} {\bibinfo {author} {\bibfnamefont {I.}~\bibnamefont
  {Booth}}, \bibinfo {author} {\bibfnamefont {R.~A.}\ \bibnamefont {Hennigar}},
  \ and\ \bibinfo {author} {\bibfnamefont {D.}~\bibnamefont {Pook-Kolb}},\
  }\href@noop {} {\  (\bibinfo {year} {2021})},\ \Eprint
  {http://arxiv.org/abs/2104.11343} {arXiv:2104.11343 [gr-qc]} \BibitemShut
  {NoStop}%
\bibitem [{\citenamefont {Ashtekar}\ \emph {et~al.}(2004)\citenamefont
  {Ashtekar}, \citenamefont {Engle}, \citenamefont {Paw{\l}owski},\ and\
  \citenamefont {Van Den~Broeck}}]{Ashtekar:2004gp}%
  \BibitemOpen
  \bibfield  {author} {\bibinfo {author} {\bibfnamefont {A.}~\bibnamefont
  {Ashtekar}}, \bibinfo {author} {\bibfnamefont {J.}~\bibnamefont {Engle}},
  \bibinfo {author} {\bibfnamefont {T.}~\bibnamefont {Paw{\l}owski}}, \ and\
  \bibinfo {author} {\bibfnamefont {C.}~\bibnamefont {Van Den~Broeck}},\ }\href
  {\doibase 10.1088/0264-9381/21/11/003} {\bibfield  {journal} {\bibinfo
  {journal} {Class. Quant. Grav.}\ }\textbf {\bibinfo {volume} {21}},\ \bibinfo
  {pages} {2549} (\bibinfo {year} {2004})},\ \Eprint
  {http://arxiv.org/abs/gr-qc/0401114} {arXiv:gr-qc/0401114} \BibitemShut
  {NoStop}%
\bibitem [{\citenamefont {Schnetter}\ \emph {et~al.}(2006)\citenamefont
  {Schnetter}, \citenamefont {Krishnan},\ and\ \citenamefont
  {Beyer}}]{Schnetter:2006yt}%
  \BibitemOpen
  \bibfield  {author} {\bibinfo {author} {\bibfnamefont {E.}~\bibnamefont
  {Schnetter}}, \bibinfo {author} {\bibfnamefont {B.}~\bibnamefont {Krishnan}},
  \ and\ \bibinfo {author} {\bibfnamefont {F.}~\bibnamefont {Beyer}},\ }\href
  {\doibase 10.1103/PhysRevD.74.024028} {\bibfield  {journal} {\bibinfo
  {journal} {Phys. Rev.}\ }\textbf {\bibinfo {volume} {D74}},\ \bibinfo {pages}
  {024028} (\bibinfo {year} {2006})},\ \Eprint
  {http://arxiv.org/abs/gr-qc/0604015} {arXiv:gr-qc/0604015} \BibitemShut
  {NoStop}%
\bibitem [{\citenamefont {Rezzolla}\ \emph {et~al.}(2010)\citenamefont
  {Rezzolla}, \citenamefont {Macedo},\ and\ \citenamefont
  {Jaramillo}}]{Rezzolla:2010df}%
  \BibitemOpen
  \bibfield  {author} {\bibinfo {author} {\bibfnamefont {L.}~\bibnamefont
  {Rezzolla}}, \bibinfo {author} {\bibfnamefont {R.~P.}\ \bibnamefont
  {Macedo}}, \ and\ \bibinfo {author} {\bibfnamefont {J.~L.}\ \bibnamefont
  {Jaramillo}},\ }\href {\doibase 10.1103/PhysRevLett.104.221101} {\bibfield
  {journal} {\bibinfo  {journal} {Phys. Rev. Lett.}\ }\textbf {\bibinfo
  {volume} {104}},\ \bibinfo {pages} {221101} (\bibinfo {year} {2010})},\
  \Eprint {http://arxiv.org/abs/1003.0873} {arXiv:1003.0873 [gr-qc]}
  \BibitemShut {NoStop}%
\bibitem [{\citenamefont {Cabero}\ and\ \citenamefont
  {Krishnan}(2015)}]{Cabero:2014nza}%
  \BibitemOpen
  \bibfield  {author} {\bibinfo {author} {\bibfnamefont {M.}~\bibnamefont
  {Cabero}}\ and\ \bibinfo {author} {\bibfnamefont {B.}~\bibnamefont
  {Krishnan}},\ }\href {\doibase 10.1088/0264-9381/32/4/045009} {\bibfield
  {journal} {\bibinfo  {journal} {Class. Quant. Grav.}\ }\textbf {\bibinfo
  {volume} {32}},\ \bibinfo {pages} {045009} (\bibinfo {year} {2015})},\
  \Eprint {http://arxiv.org/abs/1407.7656} {arXiv:1407.7656 [gr-qc]}
  \BibitemShut {NoStop}%
\bibitem [{\citenamefont {G{\"u}rlebeck}(2015)}]{Gurlebeck:2015xpa}%
  \BibitemOpen
  \bibfield  {author} {\bibinfo {author} {\bibfnamefont {N.}~\bibnamefont
  {G{\"u}rlebeck}},\ }\href {\doibase 10.1103/PhysRevLett.114.151102}
  {\bibfield  {journal} {\bibinfo  {journal} {Phys. Rev. Lett.}\ }\textbf
  {\bibinfo {volume} {114}},\ \bibinfo {pages} {151102} (\bibinfo {year}
  {2015})},\ \Eprint {http://arxiv.org/abs/1503.03240} {arXiv:1503.03240
  [gr-qc]} \BibitemShut {NoStop}%
\bibitem [{\citenamefont {Alcubierre}(2003)}]{alcubierre2003h}%
  \BibitemOpen
  \bibfield  {author} {\bibinfo {author} {\bibfnamefont {M.}~\bibnamefont
  {Alcubierre}},\ }\href {\doibase 10.1088/0264-9381/20/4/304} {\ \textbf
  {\bibinfo {volume} {20}},\ \bibinfo {pages} {607} (\bibinfo {year}
  {2003})}\BibitemShut {NoStop}%
\bibitem [{\citenamefont {Nielsen}\ \emph {et~al.}(2011)\citenamefont
  {Nielsen}, \citenamefont {Jasiulek}, \citenamefont {Krishnan},\ and\
  \citenamefont {Schnetter}}]{alex2011}%
  \BibitemOpen
  \bibfield  {author} {\bibinfo {author} {\bibfnamefont {A.~B.}\ \bibnamefont
  {Nielsen}}, \bibinfo {author} {\bibfnamefont {M.}~\bibnamefont {Jasiulek}},
  \bibinfo {author} {\bibfnamefont {B.}~\bibnamefont {Krishnan}}, \ and\
  \bibinfo {author} {\bibfnamefont {E.}~\bibnamefont {Schnetter}},\ }\href
  {\doibase 10.1103/PhysRevD.83.124022} {\bibfield  {journal} {\bibinfo
  {journal} {Phys. Rev. D}\ }\textbf {\bibinfo {volume} {83}},\ \bibinfo
  {pages} {124022} (\bibinfo {year} {2011})}\BibitemShut {NoStop}%
\bibitem [{\citenamefont {Brandt}\ and\ \citenamefont
  {Br\"ugmann}(1997)}]{PhysRevLett.78.3606}%
  \BibitemOpen
  \bibfield  {author} {\bibinfo {author} {\bibfnamefont {S.}~\bibnamefont
  {Brandt}}\ and\ \bibinfo {author} {\bibfnamefont {B.}~\bibnamefont
  {Br\"ugmann}},\ }\href {\doibase 10.1103/PhysRevLett.78.3606} {\bibfield
  {journal} {\bibinfo  {journal} {Phys. Rev. Lett.}\ }\textbf {\bibinfo
  {volume} {78}},\ \bibinfo {pages} {3606} (\bibinfo {year}
  {1997})}\BibitemShut {NoStop}%
\bibitem [{\citenamefont {Baker}\ \emph {et~al.}(2002)\citenamefont {Baker},
  \citenamefont {Campanelli}, \citenamefont {Lousto},\ and\ \citenamefont
  {Takahashi}}]{Baker:2002qf}%
  \BibitemOpen
  \bibfield  {author} {\bibinfo {author} {\bibfnamefont {J.~G.}\ \bibnamefont
  {Baker}}, \bibinfo {author} {\bibfnamefont {M.}~\bibnamefont {Campanelli}},
  \bibinfo {author} {\bibfnamefont {C.~O.}\ \bibnamefont {Lousto}}, \ and\
  \bibinfo {author} {\bibfnamefont {R.}~\bibnamefont {Takahashi}},\ }\href
  {\doibase 10.1103/PhysRevD.65.124012} {\bibfield  {journal} {\bibinfo
  {journal} {Phys. Rev.}\ }\textbf {\bibinfo {volume} {D65}},\ \bibinfo {pages}
  {124012} (\bibinfo {year} {2002})},\ \Eprint
  {http://arxiv.org/abs/astro-ph/0202469} {arXiv:astro-ph/0202469 [astro-ph]}
  \BibitemShut {NoStop}%
\bibitem [{\citenamefont {Dreyer}\ \emph {et~al.}(2003)\citenamefont {Dreyer},
  \citenamefont {Krishnan}, \citenamefont {Shoemaker},\ and\ \citenamefont
  {Schnetter}}]{Dreyer:2002mx}%
  \BibitemOpen
  \bibfield  {author} {\bibinfo {author} {\bibfnamefont {O.}~\bibnamefont
  {Dreyer}}, \bibinfo {author} {\bibfnamefont {B.}~\bibnamefont {Krishnan}},
  \bibinfo {author} {\bibfnamefont {D.}~\bibnamefont {Shoemaker}}, \ and\
  \bibinfo {author} {\bibfnamefont {E.}~\bibnamefont {Schnetter}},\ }\href
  {\doibase 10.1103/PhysRevD.67.024018} {\bibfield  {journal} {\bibinfo
  {journal} {Phys. Rev.}\ }\textbf {\bibinfo {volume} {D67}},\ \bibinfo {pages}
  {024018} (\bibinfo {year} {2003})},\ \Eprint
  {http://arxiv.org/abs/gr-qc/0206008} {arXiv:gr-qc/0206008} \BibitemShut
  {NoStop}%
\bibitem [{\citenamefont {L{\"{o}}ffler}\ \emph {et~al.}(2012)\citenamefont
  {L{\"{o}}ffler}, \citenamefont {Faber}, \citenamefont {Bentivegna},
  \citenamefont {Bode}, \citenamefont {Diener}, \citenamefont {Haas},
  \citenamefont {Hinder}, \citenamefont {Mundim}, \citenamefont {Ott},
  \citenamefont {Schnetter}, \citenamefont {Allen}, \citenamefont
  {Campanelli},\ and\ \citenamefont {Laguna}}]{Loffler:2011ay}%
  \BibitemOpen
  \bibfield  {author} {\bibinfo {author} {\bibfnamefont {F.}~\bibnamefont
  {L{\"{o}}ffler}}, \bibinfo {author} {\bibfnamefont {J.}~\bibnamefont
  {Faber}}, \bibinfo {author} {\bibfnamefont {E.}~\bibnamefont {Bentivegna}},
  \bibinfo {author} {\bibfnamefont {T.}~\bibnamefont {Bode}}, \bibinfo {author}
  {\bibfnamefont {P.}~\bibnamefont {Diener}}, \bibinfo {author} {\bibfnamefont
  {R.}~\bibnamefont {Haas}}, \bibinfo {author} {\bibfnamefont {I.}~\bibnamefont
  {Hinder}}, \bibinfo {author} {\bibfnamefont {B.~C.}\ \bibnamefont {Mundim}},
  \bibinfo {author} {\bibfnamefont {C.~D.}\ \bibnamefont {Ott}}, \bibinfo
  {author} {\bibfnamefont {E.}~\bibnamefont {Schnetter}}, \bibinfo {author}
  {\bibfnamefont {G.}~\bibnamefont {Allen}}, \bibinfo {author} {\bibfnamefont
  {M.}~\bibnamefont {Campanelli}}, \ and\ \bibinfo {author} {\bibfnamefont
  {P.}~\bibnamefont {Laguna}},\ }\href {\doibase
  doi:10.1088/0264-9381/29/11/115001} {\bibfield  {journal} {\bibinfo
  {journal} {Class. Quantum Grav.}\ }\textbf {\bibinfo {volume} {29}},\
  \bibinfo {pages} {115001} (\bibinfo {year} {2012})},\ \Eprint
  {http://arxiv.org/abs/arXiv:1111.3344 [gr-qc]} {arXiv:1111.3344 [gr-qc]}
  \BibitemShut {NoStop}%
\bibitem [{EinsteinToolkit()}]{EinsteinToolkit:web}%
  \BibitemOpen
  EinsteinToolkit,\ \href@noop {} {\enquote {\bibinfo {title} {{Einstein
  Toolkit}: Open software for relativistic astrophysics},}\ }\bibinfo {note}
  {\url{http://einsteintoolkit.org/}}\BibitemShut {NoStop}%
\bibitem [{\citenamefont {Ansorg}\ \emph {et~al.}(2004)\citenamefont {Ansorg},
  \citenamefont {Br{\"u}gmann},\ and\ \citenamefont {Tichy}}]{Ansorg:2004ds}%
  \BibitemOpen
  \bibfield  {author} {\bibinfo {author} {\bibfnamefont {M.}~\bibnamefont
  {Ansorg}}, \bibinfo {author} {\bibfnamefont {B.}~\bibnamefont
  {Br{\"u}gmann}}, \ and\ \bibinfo {author} {\bibfnamefont {W.}~\bibnamefont
  {Tichy}},\ }\href {\doibase 10.1103/PhysRevD.70.064011} {\bibfield  {journal}
  {\bibinfo  {journal} {Phys. Rev. D}\ }\textbf {\bibinfo {volume} {70}},\
  \bibinfo {pages} {064011} (\bibinfo {year} {2004})},\ \Eprint
  {http://arxiv.org/abs/arXiv:gr-qc/0404056} {arXiv:gr-qc/0404056} \BibitemShut
  {NoStop}%
\bibitem [{\citenamefont {Alcubierre}\ \emph {et~al.}(2000)\citenamefont
  {Alcubierre}, \citenamefont {Allen}, \citenamefont {Br{\"u}gmann},
  \citenamefont {Dramlitsch}, \citenamefont {Font}, \citenamefont
  {Papadopoulos}, \citenamefont {Seidel}, \citenamefont {Stergioulas},
  \citenamefont {Suen},\ and\ \citenamefont {Takahashi}}]{Alcubierre:2000xu}%
  \BibitemOpen
  \bibfield  {author} {\bibinfo {author} {\bibfnamefont {M.}~\bibnamefont
  {Alcubierre}}, \bibinfo {author} {\bibfnamefont {G.}~\bibnamefont {Allen}},
  \bibinfo {author} {\bibfnamefont {B.}~\bibnamefont {Br{\"u}gmann}}, \bibinfo
  {author} {\bibfnamefont {T.}~\bibnamefont {Dramlitsch}}, \bibinfo {author}
  {\bibfnamefont {J.~A.}\ \bibnamefont {Font}}, \bibinfo {author}
  {\bibfnamefont {P.}~\bibnamefont {Papadopoulos}}, \bibinfo {author}
  {\bibfnamefont {E.}~\bibnamefont {Seidel}}, \bibinfo {author} {\bibfnamefont
  {N.}~\bibnamefont {Stergioulas}}, \bibinfo {author} {\bibfnamefont {W.-M.}\
  \bibnamefont {Suen}}, \ and\ \bibinfo {author} {\bibfnamefont
  {R.}~\bibnamefont {Takahashi}},\ }\href {\doibase 10.1103/PhysRevD.62.044034}
  {\bibfield  {journal} {\bibinfo  {journal} {Phys. Rev.}\ }\textbf {\bibinfo
  {volume} {D62}},\ \bibinfo {pages} {044034} (\bibinfo {year} {2000})},\
  \Eprint {http://arxiv.org/abs/gr-qc/0003071} {arXiv:gr-qc/0003071 [gr-qc]}
  \BibitemShut {NoStop}%
\bibitem [{\citenamefont {Alcubierre}\ \emph {et~al.}(2003)\citenamefont
  {Alcubierre}, \citenamefont {Br{\"u}gmann}, \citenamefont {Diener},
  \citenamefont {Koppitz}, \citenamefont {Pollney}, \citenamefont {Seidel},\
  and\ \citenamefont {Takahashi}}]{Alcubierre:2002kk}%
  \BibitemOpen
  \bibfield  {author} {\bibinfo {author} {\bibfnamefont {M.}~\bibnamefont
  {Alcubierre}}, \bibinfo {author} {\bibfnamefont {B.}~\bibnamefont
  {Br{\"u}gmann}}, \bibinfo {author} {\bibfnamefont {P.}~\bibnamefont
  {Diener}}, \bibinfo {author} {\bibfnamefont {M.}~\bibnamefont {Koppitz}},
  \bibinfo {author} {\bibfnamefont {D.}~\bibnamefont {Pollney}}, \bibinfo
  {author} {\bibfnamefont {E.}~\bibnamefont {Seidel}}, \ and\ \bibinfo {author}
  {\bibfnamefont {R.}~\bibnamefont {Takahashi}},\ }\href {\doibase
  10.1103/PhysRevD.67.084023} {\bibfield  {journal} {\bibinfo  {journal} {Phys.
  Rev.}\ }\textbf {\bibinfo {volume} {D67}},\ \bibinfo {pages} {084023}
  (\bibinfo {year} {2003})},\ \Eprint {http://arxiv.org/abs/gr-qc/0206072}
  {arXiv:gr-qc/0206072 [gr-qc]} \BibitemShut {NoStop}%
\bibitem [{\citenamefont {Brown}\ \emph {et~al.}(2009)\citenamefont {Brown},
  \citenamefont {Diener}, \citenamefont {Sarbach}, \citenamefont {Schnetter},\
  and\ \citenamefont {Tiglio}}]{Brown:2008sb}%
  \BibitemOpen
  \bibfield  {author} {\bibinfo {author} {\bibfnamefont {J.~D.}\ \bibnamefont
  {Brown}}, \bibinfo {author} {\bibfnamefont {P.}~\bibnamefont {Diener}},
  \bibinfo {author} {\bibfnamefont {O.}~\bibnamefont {Sarbach}}, \bibinfo
  {author} {\bibfnamefont {E.}~\bibnamefont {Schnetter}}, \ and\ \bibinfo
  {author} {\bibfnamefont {M.}~\bibnamefont {Tiglio}},\ }\href {\doibase
  10.1103/PhysRevD.79.044023} {\bibfield  {journal} {\bibinfo  {journal} {Phys.
  Rev. D}\ }\textbf {\bibinfo {volume} {79}},\ \bibinfo {pages} {044023}
  (\bibinfo {year} {2009})},\ \Eprint {http://arxiv.org/abs/arXiv:0809.3533
  [gr-qc]} {arXiv:0809.3533 [gr-qc]} \BibitemShut {NoStop}%
\bibitem [{\citenamefont {Pollney}\ \emph {et~al.}(2011)\citenamefont
  {Pollney}, \citenamefont {Reisswig}, \citenamefont {Schnetter}, \citenamefont
  {Dorband},\ and\ \citenamefont {Diener}}]{PhysRevD.83.044045}%
  \BibitemOpen
  \bibfield  {author} {\bibinfo {author} {\bibfnamefont {D.}~\bibnamefont
  {Pollney}}, \bibinfo {author} {\bibfnamefont {C.}~\bibnamefont {Reisswig}},
  \bibinfo {author} {\bibfnamefont {E.}~\bibnamefont {Schnetter}}, \bibinfo
  {author} {\bibfnamefont {N.}~\bibnamefont {Dorband}}, \ and\ \bibinfo
  {author} {\bibfnamefont {P.}~\bibnamefont {Diener}},\ }\href {\doibase
  10.1103/PhysRevD.83.044045} {\bibfield  {journal} {\bibinfo  {journal} {Phys.
  Rev. D}\ }\textbf {\bibinfo {volume} {83}},\ \bibinfo {pages} {044045}
  (\bibinfo {year} {2011})}\BibitemShut {NoStop}%
\bibitem [{\citenamefont {Thornburg}(1996)}]{Thornburg:1995cp}%
  \BibitemOpen
  \bibfield  {author} {\bibinfo {author} {\bibfnamefont {J.}~\bibnamefont
  {Thornburg}},\ }\href {\doibase 10.1103/PhysRevD.54.4899} {\bibfield
  {journal} {\bibinfo  {journal} {Phys. Rev. D}\ }\textbf {\bibinfo {volume}
  {54}},\ \bibinfo {pages} {4899} (\bibinfo {year} {1996})},\ \Eprint
  {http://arxiv.org/abs/arXiv:gr-qc/9508014} {arXiv:gr-qc/9508014} \BibitemShut
  {NoStop}%
\bibitem [{\citenamefont {Thornburg}(2004)}]{Thornburg:2003sf}%
  \BibitemOpen
  \bibfield  {author} {\bibinfo {author} {\bibfnamefont {J.}~\bibnamefont
  {Thornburg}},\ }\href {\doibase 10.1088/0264-9381/21/2/026} {\bibfield
  {journal} {\bibinfo  {journal} {Class. Quant. Grav.}\ }\textbf {\bibinfo
  {volume} {21}},\ \bibinfo {pages} {743} (\bibinfo {year} {2004})},\ \Eprint
  {http://arxiv.org/abs/gr-qc/0306056} {arXiv:gr-qc/0306056} \BibitemShut
  {NoStop}%
\bibitem [{\citenamefont {Wardell}\ \emph {et~al.}(2016)\citenamefont
  {Wardell}, \citenamefont {Hinder},\ and\ \citenamefont
  {Bentivegna}}]{wardell_barry_2016_155394}%
  \BibitemOpen
  \bibfield  {author} {\bibinfo {author} {\bibfnamefont {B.}~\bibnamefont
  {Wardell}}, \bibinfo {author} {\bibfnamefont {I.}~\bibnamefont {Hinder}}, \
  and\ \bibinfo {author} {\bibfnamefont {E.}~\bibnamefont {Bentivegna}},\
  }\href {\doibase 10.5281/zenodo.155394} {\enquote {\bibinfo {title}
  {{Simulation of GW150914 binary black hole merger using the Einstein
  Toolkit}},}\ } (\bibinfo {year} {2016}),\ \bibinfo {note}
  {\url{https://doi.org/10.5281/zenodo.155394}}\BibitemShut {NoStop}%
\bibitem [{\citenamefont {Healy}\ \emph {et~al.}(2014)\citenamefont {Healy},
  \citenamefont {Lousto},\ and\ \citenamefont {Zlochower}}]{Healy:2014yta}%
  \BibitemOpen
  \bibfield  {author} {\bibinfo {author} {\bibfnamefont {J.}~\bibnamefont
  {Healy}}, \bibinfo {author} {\bibfnamefont {C.~O.}\ \bibnamefont {Lousto}}, \
  and\ \bibinfo {author} {\bibfnamefont {Y.}~\bibnamefont {Zlochower}},\ }\href
  {\doibase 10.1103/PhysRevD.90.104004} {\bibfield  {journal} {\bibinfo
  {journal} {Phys. Rev.}\ }\textbf {\bibinfo {volume} {D90}},\ \bibinfo {pages}
  {104004} (\bibinfo {year} {2014})},\ \Eprint {http://arxiv.org/abs/1406.7295}
  {arXiv:1406.7295 [gr-qc]} \BibitemShut {NoStop}%
\bibitem [{\citenamefont {Healy}\ and\ \citenamefont
  {Lousto}(2017)}]{PhysRevD.95.024037}%
  \BibitemOpen
  \bibfield  {author} {\bibinfo {author} {\bibfnamefont {J.}~\bibnamefont
  {Healy}}\ and\ \bibinfo {author} {\bibfnamefont {C.~O.}\ \bibnamefont
  {Lousto}},\ }\href {\doibase 10.1103/PhysRevD.95.024037} {\bibfield
  {journal} {\bibinfo  {journal} {Phys. Rev. D}\ }\textbf {\bibinfo {volume}
  {95}},\ \bibinfo {pages} {024037} (\bibinfo {year} {2017})}\BibitemShut
  {NoStop}%
\bibitem [{RITcatalog()}]{RITcatalog:web}%
  \BibitemOpen
  RITcatalog,\ \href@noop {} {\enquote {\bibinfo {title} {{RIT Catalog for
  Numerical Simulations}},}\ }\bibinfo {note}
  {\url{https://ccrg.rit.edu/~RITCatalog/}}\BibitemShut {NoStop}%
\bibitem [{\citenamefont {Xu}\ and\ \citenamefont {Paik}(2016)}]{xupaik2016}%
  \BibitemOpen
  \bibfield  {author} {\bibinfo {author} {\bibfnamefont {P.}~\bibnamefont
  {Xu}}\ and\ \bibinfo {author} {\bibfnamefont {H.~J.}\ \bibnamefont {Paik}},\
  }\href {\doibase 10.1103/PhysRevD.93.044057} {\bibfield  {journal} {\bibinfo
  {journal} {Phys. Rev. D}\ }\textbf {\bibinfo {volume} {93}},\ \bibinfo
  {pages} {044057} (\bibinfo {year} {2016})}\BibitemShut {NoStop}%
\bibitem [{\citenamefont {Thornburg}(2003)}]{Thornburg_2003}%
  \BibitemOpen
  \bibfield  {author} {\bibinfo {author} {\bibfnamefont {J.}~\bibnamefont
  {Thornburg}},\ }\href {\doibase 10.1088/0264-9381/21/2/026} {\bibfield
  {journal} {\bibinfo  {journal} {Classical and Quantum Gravity}\ }\textbf
  {\bibinfo {volume} {21}},\ \bibinfo {pages} {743} (\bibinfo {year}
  {2003})}\BibitemShut {NoStop}%
\bibitem [{\citenamefont {Choustikov}(2020)}]{etk_student_guide_2020}%
  \BibitemOpen
  \bibfield  {author} {\bibinfo {author} {\bibfnamefont {N.}~\bibnamefont
  {Choustikov}},\ }\href@noop {} {\enquote {\bibinfo {title} {The einstein
  toolkit: A student's guide},}\ } (\bibinfo {year} {2020}),\ \Eprint
  {http://arxiv.org/abs/2011.13314} {arXiv:2011.13314 [gr-qc]} \BibitemShut
  {NoStop}%
\end{thebibliography}%
\end{document}